\documentclass[useAMS,usenatbib]{mn2e}
\usepackage{graphicx,rotate,url,mathptmx,lscape,times,array,verbatim,caption,float,verbatim,moreverb}
\pdfoutput=1
\newcommand{\chia}{Chianti}
\newcommand{\op}{Opacity Project}
\newcommand{\kz}{Kurucz}
\usepackage{./common/aas_macros}
\usepackage{color}
\newcommand{\Cloudy}{\textsc{Cloudy}}

\font\manual=manfnt at 7pt \def\dbend{\hbox{\raise0.9ex\hbox{\manual\char127\hspace{0.6em}}}}

\providecommand{\e}[1]{\ensuremath{\times 10^{#1}}}

\newcounter{INTERNALionstage}

\def\gtsim{\mathrel{\hbox{\rlap{\hbox{\lower4pt\hbox{$\sim$}}}\hbox{$>$}}}}
\def\lesssim{\mathrel{\hbox{\rlap{\hbox{\lower4pt\hbox{$\sim$}}}\hbox{$<$}}}}

\def\A{{\rm\thinspace \AA}}

\def\K{{\rm\thinspace K}}

\def\pcc{{\rm\thinspace cm^{-3}}}

\def\fei{\mbox{{\rm Fe~{\sc i}}}}
\def\feii{\mbox{{\rm Fe~{\sc ii}}}}
\def\feiii{\mbox{{\rm Fe~{\sc iii}}}}
\def\feiv{\mbox{{\rm Fe~{\sc iv}}}}
\def\fev{\mbox{{\rm Fe~{\sc v}}}}

\def\fevii{\mbox{{\rm Fe~{\sc vii}}}}
\def\feviii{\mbox{{\rm Fe~{\sc viii}}}}

\def\fexix{\mbox{{\rm Fe~{\sc xix}}}}

\def\fexxiv{\mbox{{\rm Fe~{\sc xxiv}}}}
\def\fexxv{\mbox{{\rm Fe~{\sc xxv}}}}
\def\fexxvi{\mbox{{\rm Fe~{\sc xxvi}}}}
\def\hplus{\mbox{{\rm H}$^+$}}

\def\hO{\mbox{{\rm H}$^0$}}

\def\heo{\mbox{{\rm He}$^0$}}
\def\heplus{\mbox{{\rm He}$^+$}}
\def\hePP{\mbox{{\rm He}$^{2+}$}}
\DeclareMathAlphabet{\vib}{OML}{cmm}{m}{it}
\title[Radiative Cooling]{Radiative cooling in collisionally and photo ionized plasmas
\thanks{Contains material \copyright\ British Crown copyright 2011/MoD}}

\author[M. L. Lykins, et al.]
	{\parbox[]{6.0in} 
	{
       M. L. Lykins,$^{1}$
	G. J. Ferland,$^{1}$		
	Ryan L. Porter,$^{2}$\\
	Peter A. M. van Hoof,$^{3}$
	R.J.R. Williams,$^{4}$ and
	Orly Gnat$^{5}$
\\		
 \footnotesize
$^1$University of Kentucky, Lexington, KY 40506, USA \\
$^2$Department of Physics and Astronomy and Center for Simulational Physics, University of Georgia, Athens, GA 30602, USA\\
$^3$Royal Observatory of Belgium, Ringlaan 3, 1180 Brussels, Belgium\\
$^4$AWE plc, Aldermaston, Reading RG7 4PR, UK\\
$^5$Racah Institute of Physics, The Hebrew University, Jerusalem 91904, Israel\\
	}
}

\pagerange{\pageref{firstpage}--\pageref{lastpage}}
\pubyear{} 

\begin{document}

\maketitle

\label{firstpage}

\begin{abstract}
%\noindent
We discuss recent improvements in the calculation of the radiative cooling
in both collisionally and photo ionized plasmas.
We are extending the spectral simulation code Cloudy
so that as much as possible of the underlying atomic data is taken from
external databases, some created by others, some developed by the
Cloudy team.
This paper focuses on recent changes in the treatment of many
stages of ionization of iron, and discusses its extensions to 
other elements.  
The H-like and He-like ions are treated in the iso-electronic approach described
previously.  
\feii\ is a special case treated with a large model atom.
Here we focus on \feiii\ through \fexxiv, ions which are important contributors to the radiative cooling
of hot ($T\sim 10^5 - 10^7 \K$) plasmas and for X-ray spectroscopy.
We use the Chianti atomic database to greatly expand the
number of transitions in the cooling function.
Chianti only includes lines that have atomic data computed by
sophisticated methods.
This limits the line list to lower excitation, longer wavelength, transitions.
We had previously included lines from the Opacity Project database, 
which tends to include higher energy, shorter wavelength, transitions.
These were combined with various forms of the ``g-bar'' approximation,
a highly approximate method of estimating collision rates.
For several iron ions the two databases are almost entirely complementary.
We adopt a hybrid approach in which we use Chianti where possible, 
supplemented by
lines from the Opacity Project for shorter wavelength transitions.
The total cooling including the lightest thirty element differs from some previous 
calculations by significant amounts.
\end{abstract}

\begin{keywords}
TBD
\end{keywords}

\section{Introduction}

This paper describes recent advances in the treatment of
cooling in the spectral simulation code \Cloudy. 
The companion paper, Williams et al.\ (in preparation), determines
the emission spectrum of a non-equilibrium cooling and recombining
plasma where the physics is largely driven by radiative cooling.

Cloudy performs, as its primary goal, a full simulation
of the microphysics of a non-equilibrium gas.  
As described in the last major review,
\citep{FerlandEtAl98}, the code is designed
to incorporate the essential microphysics of
gas between the molecular and fully ionized limits,
with densities between LTE and the low-density limit,
and with temperatures between the current CMB and $10^{10} \K$.
\citet{AGN3}, hereafter AGN3, provides many details of this physics.
Our approach is to treat the microphysics in great detail, using
basic cross sections and transition rates where possible, to do
exactly what nature does under this broad range of conditions.

The atomic / molecular database is the essential difficulty
in producing a full simulation of a non-equilibrium gas.
Atomic data, like the underlying quantum mechanics, is complex
due to the idiosyncrasies that are characteristic of each molecule or ion.
Through much of its history we have added physical
processes to the code as special cases, each treated
individually.
A large model of the \feii\ atom was developed by
Katya Verner as part of her PhD thesis \citep{Verner1999},
while Gargi Shaw did a complete model of the hydrogen
molecule as part of her PhD thesis \citep{Shaw2005}.
Ryan Porter developed a unified treatment of the He-like iso-sequence
[\citet{Porter2005} and \citet{PorterFerland2007}]
and extended it to include the H-like sequence
\citep{LuridianaEtAl09} as part of his thesis.

Additional contributors to the emission spectrum and cooling
were added on a line by line basis. 
Initially a range of lines based on previous calculations
of the cooling function were used
(\citet{Kato1976}; \citet{Gaetz1983}).
Additional lines were added on an ad hoc basis.
Finally, all Opacity Project \citep{Seaton1987} permitted lines
that directly connect to the ground state were added 
(see also \citet{VernerVerner1996}).
The Opacity Project did not compute collision rates
so the emission data were combined with various forms
of the ``g-bar'' approximation, an approximate relationship
between the collision rate and other atomic parameters,
to compute the emission.
The current implementation uses g-bar approximations
from \citet{Mewe1972}, \citet{Gaetz1983}, and
\citet{MeweEtAl85}.
Additionally, level energies and line wavelengths for OP lines 
are uncertain by roughly 15\%,
although these energies can be removed by comparing with experiments, 
as was done by 
\citet{VernerVerner1996}. 
The new improvements are described next.

\section{Calculations}

We have updated \Cloudy\ to use iron lines from the \chia \footnote{CHIANTI is a collaborative project involving the NRL (USA), the Universities of Florence (Italy) and Cambridge (UK), and George Mason University (USA).} database 
(\cite{Dere.K97CHIANTI---an-atomic-database-for-emission}; \cite{Landi2012}) version 7.
Of the astrophysically abundant elements, iron is the one
with the richest spectrum and an element which has been
an emphasis for the \chia\ project.
\Cloudy\ will now make use of the experimentally measured lines from the \chia\ database for \feiv\ through \fexxiv.
The \chia\ implementations of \fexxv\ and \fexxvi, He-like and H-like iron, are not 
used because 
we treat these using the iso-electronic approach described in \citet{Porter2005}, \citet{PorterFerland2007}, and \citet{LuridianaEtAl09}.
The special case of \feii\ continues to be treated with the Verner model atom.

In situations where \chia\ has provided transitions without collision strengths, the g-bar approximation from \citet{Mewe1972} is used.
The transitions are estimated to be allowed or forbidden based on their oscillator strengths values ($gf$).
Transitions where $gf \geq 1\e{-8}$ are classified as allowed, all others are forbidden.
After classification the appropriate g-bar approximation equation is used.

In addition to the \chia\ database lines for \feiv\ through \fexxiv, we added \feiii\ lines from the \kz\ Atomic Database \citep{Kurucz2009}.
The \kz\ lines, like the \op\ lines, lack collisional data so we use the g-bar approximation.
We will use the \chia\ and \kz\ data where they are available since they 
have more accurate energies.
We supplement these with \op\ lines that come from levels that have higher excitation than
those in  \chia\ and \kz. This implementation, called the Hybrid configuration,  
gives \Cloudy\ the greatest accuracy and wavelength coverage.

In remainder of this section we outline our approach and compute cooling functions
for collisionally and photo ionized plasmas.
We find surprisingly good agreement with older calculations, and with others
based on a detailed incorporation of the atomic physics, but not with several recent studies.

A calculation of the gas cooling involves several steps.  
First the ionization or chemical state of the gas must be determined.
This distribution is then used to compute the cooling, the rate that collisions convert
kinetic energy into light.
The following sections give details concerning these calculations.

\subsection{The ionization balance}

This paper is limited to atomic and ionic cooling, and so is limited
to temperatures greater than $10^4$ K.  
A calculation of the ionization balance involves rates for 
collisional and photo ionization,
and various recombination processes.  
These are described in the following subsections.

\subsubsection{Collisional Ionization}

\Cloudy\ has used the collisional ionization rate coefficients tabulated by \citet{Voronov1997} 
since soon after the publication of that paper.
More recently \citet{Dere.K07Ionization-rate-coefficients-for-the-elements}
presented a new compilation which is largely in excellent agreement with \citet{Voronov1997}.
The \citet{Dere.K07Ionization-rate-coefficients-for-the-elements} recommendations originated 
with experiments from different sources and theoretical calculations using 
the Flexible Atomic Code (FAC) described in \citet{GuFAC2002}.  
The collisional ionization rate coefficients of  
\citet{Dere.K07Ionization-rate-coefficients-for-the-elements} and \citet{Voronov1997} 
only differ significantly for about half a dozen ions.
We provided options which will allow \Cloudy\ to use either set of rates.

We have implemented these data in the following way in our default calculation. 
The \citet{Dere.K07Ionization-rate-coefficients-for-the-elements} coefficients are provided in a 
discrete format for selected temperatures which do not span the temperature range
needed by \Cloudy.
\citet{Voronov1997} provides continuous functions which are valid for any temperature,
going to the appropriate low and high temperature limits. 
We scaled the \citet{Voronov1997} rates to the values of 
\citet{Dere.K07Ionization-rate-coefficients-for-the-elements}. 
For each species, the scale factor is the 
ratio of the \citet{Dere.K07Ionization-rate-coefficients-for-the-elements} to \citet{Voronov1997} rates 
at the center of the temperature range where the ion abundance peaks.
These scaling coefficients are typically within 10\% of unity.
This is now the default for \Cloudy.

\begin{table}
  \centering
  \caption{Log of fractional ionization of hydrogen and helium. See the online version of this table which includes the 30 lightest elements and has higher temperature resolution.}
\label{tab:iondist}
    \begin{tabular}{cccccc}
\hline
\hline
    $T_e (K)$   & \hO   & \hplus   & \heo  & \heplus  & \hePP \\
\hline
    4    & 0.00 & -3.34 & 0.00 & --   & -- \\
    4.5  & -2.56 & 0.00 & -0.55 & -0.15 & -5.71 \\
    5    & -4.71 & 0.00 & -4.15 & -0.95 & -0.05 \\
    5.5  & -5.84 & 0.00 & -7.17 & -3.35 & 0.00 \\
    6    & -6.59 & 0.00 & --   & -4.53 & 0.00 \\
    6.5  & -7.16 & 0.00 & --   & -5.30 & 0.00 \\
    7    & -7.68 & 0.00 & --   & -5.89 & 0.00 \\
    7.5  & -8.18 & 0.00 & --   & -6.43 & 0.00 \\
    8    & -8.70 & 0.00 & --   & -6.98 & 0.00 \\
    8.5  & --   & 0.00 & --   & -7.55 & 0.00 \\
    9    & --   & 0.00 & --   & -8.14 & 0.00 \\
\hline
    \end{tabular}%
\end{table}%

\subsubsection{Photoionization}
The photoionization cross section database 
remains unchanged from \citet{FerlandEtAl98}.

\subsubsection{Recombination coefficients}
\label{recombinationcoefficients}
\footnotetext[2]{Badnell site: http://amdpp.phys.strath.ac.uk/tamoc/DATA/}

We updated \Cloudy\ with the latest radiative (RR) and dielectronic recombination (DR) rate coefficients from Badnell's website\footnotemark\ (\cite{BadnellEtAl03} and \cite{Badnell06}). 
The update to the DR rate coefficients 
includes recent data for the argon-like isoelectronic sequence \citep{Badnell2010}
and \citet{Abdel-Naby.S12Dielectronic-recombination-data-for-dynamic} for 
the aluminium-like sequence. 
We use an ion-specific ``mean'' value for species which are not covered by
the Badnell database, as was described by 
\citet{Ali.B91The-NE-III-O-II-forbidden-line-spectrum-as-an-ionization}. 

\subsection{Temperatures of peak abundance for photoionization and collisional ionization}
\label{peakabundance}

Our goal is to simulate both photo and collisionally ionized plasmas, for a wide range
of chemical abundances and energy sources.
There are two limiting cases that are considered in much of the active literature.
In the photoionization case the gas is irradiated by an external energy source
and the equations of thermal and ionization equilibrium are solved \citep{AGN3}.
The gas kinetic temperature will depend on both the spectral energy distribution (SED) and the composition, being higher
for harder SEDs or lower abundances.
The ionization distribution is set by the balance between photoionization
and recombination rates, and is not directly set by the kinetic temperature.
Similarly, the cooling is not a unique function of the temperature in this case.
In the collisional ionization case the gas kinetic temperature is often specified,
having been set by physics external to the problem, although it would be
possible to specify a heating rate and determine the temperature.
The ionization distribution is set by the balance between collisional ionization
and recombination rates.
The ionization and cooling are directly determined by the temperature in this
collisional case.

\begin{figure}
\includegraphics[width=\columnwidth, keepaspectratio]{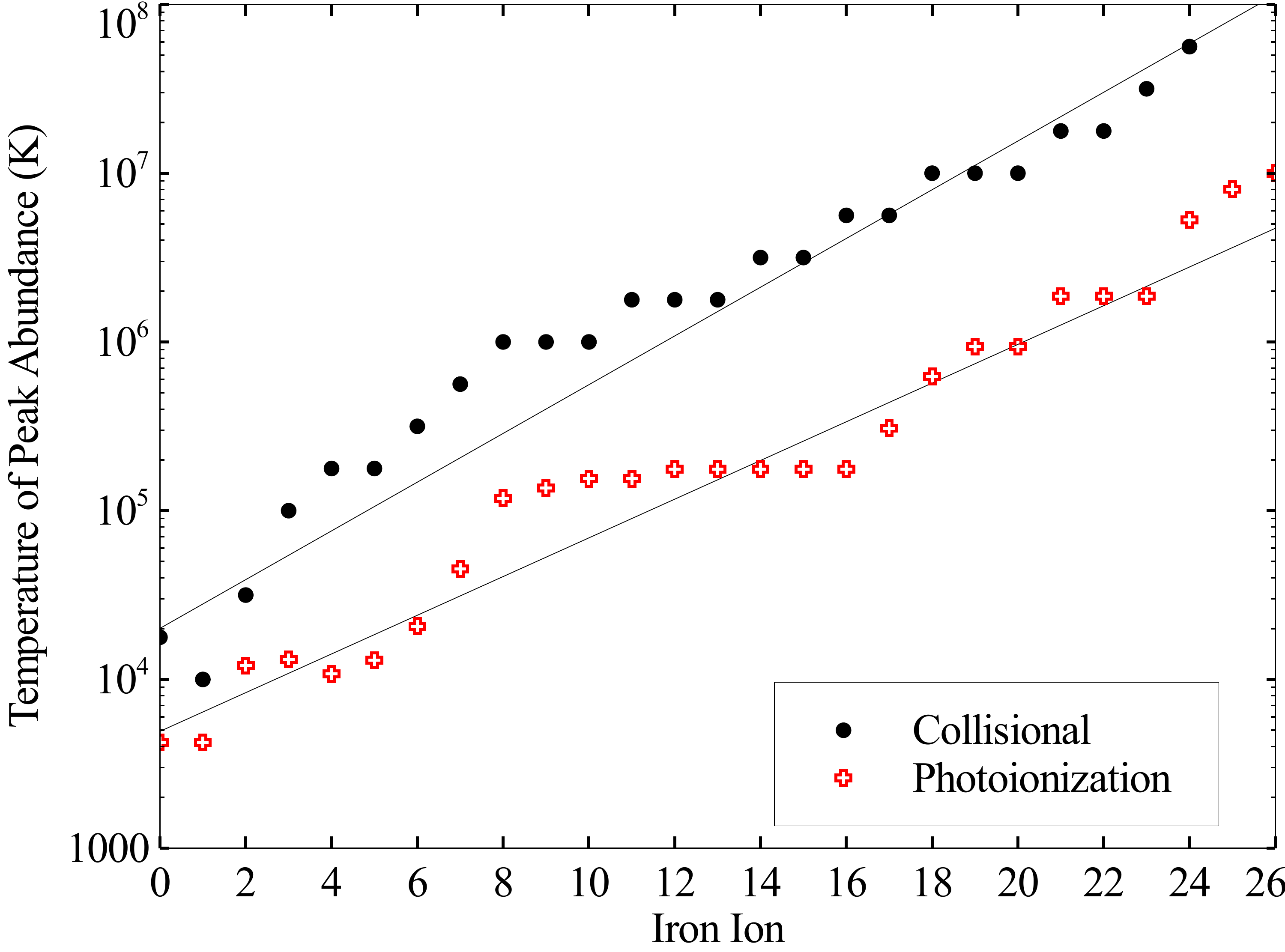}
\caption{Temperature of peak abundance per iron ion for the collisional
and photo ionization cases.  The gas is cooler in photoionization equilibrium,
which affects the strategy used to compute the cooling rate.
}
\label{fig:tmaxvsion}
\end{figure}

The temperature where a particular ion reaches it peak abundance is different for
these two cases.
We computed a series of models using solar abundances and a density of $1 \pcc$.
Given these assumptions the gas ionization, for an optically thin cell, depends on
the gas temperature in the collisional case, and on the intensity of light striking the
gas in the photoionization case.
The ``ionization parameter'', a way of specifying the intensity of the radiation field,
was varied in the photoionization case and
the \citet{MathewsFerland87} SED of a typical Active Galactic Nucleus was used.
In the collisional case the kinetic temperature was varied and
the ionization balance determined.
The temperature where each successive iron ion peaked was then determined,
and is plotted in Figure \ref{fig:tmaxvsion}.
Figure \ref{fig:tmaxvsion} shows that for a given iron ion, the temperature of peak abundance is significantly higher when collisions rather than photons dominate ionizations.

This has two effects on the calculation of the gas cooling.
In the collisional ionization case the gas kinetic temperature is not much lower
than the ionization energy of the ion.  
This means that very highly excited levels can be populated by thermal collisions.
In the photoionization case the temperature is significantly lower, meaning the
the cooling will be dominated by a few lower levels.
This affects our strategy in optimizing our selection of the number of levels to include
in atomic models.

\subsection{Line Cooling}
\label{linecooling}

\subsubsection{Cloudy Hybrid}
\label{cloudyhybrid}

\begin{figure}
\includegraphics[width=\columnwidth, keepaspectratio]{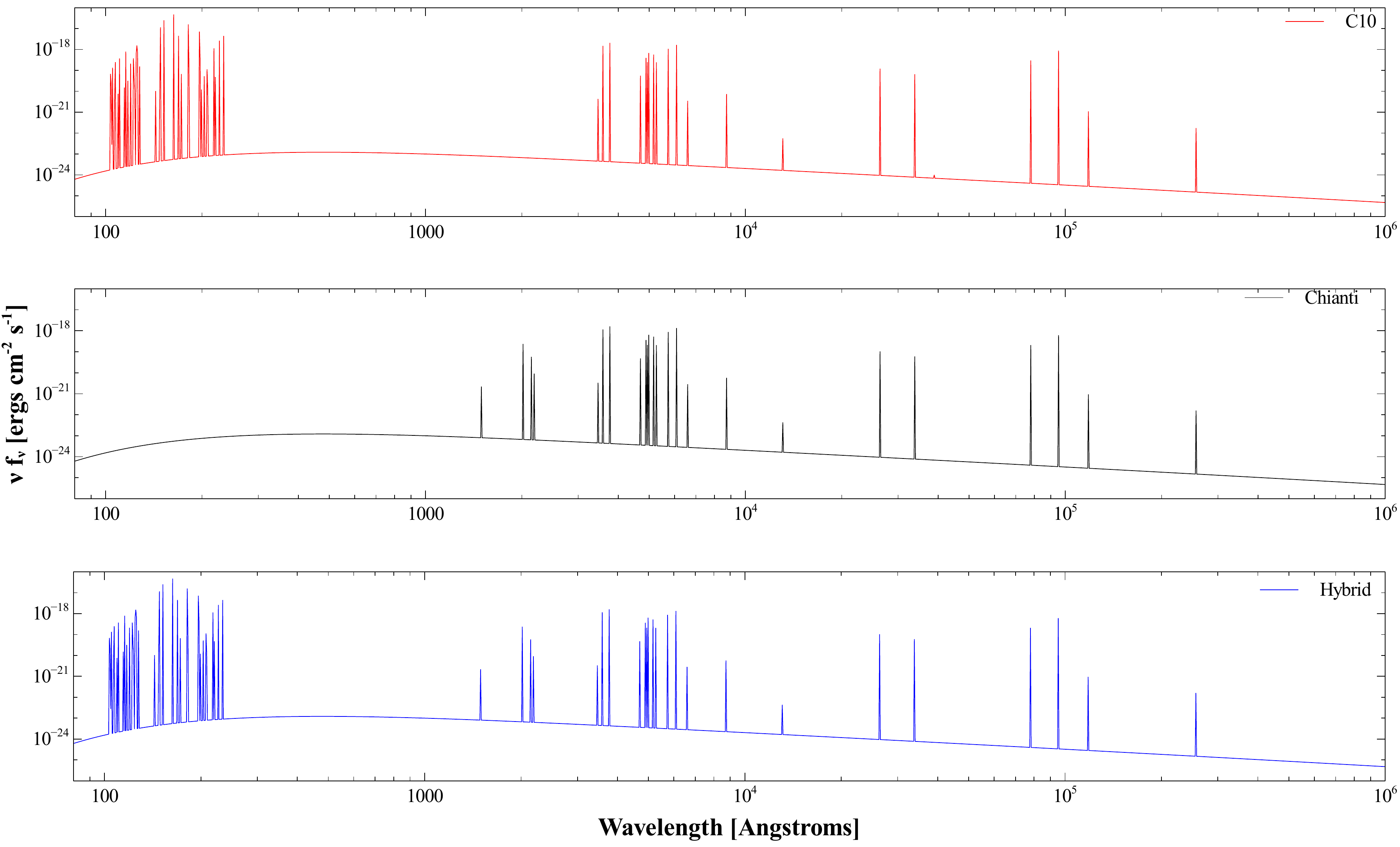}
\caption{Plots of \fevii\ spectra for a collisionally ionized gas under three different \Cloudy\ configurations; 
\op\ with the \Cloudy\ internal database (labelled as C10), \chia\ only, and our new default Hybrid configuration (\chia\ + \op).
Spectra for other iron ions considered in this paper are shown in
the on-line material.}
\label{fig:3spectra}
\end{figure}

We have long included all resonance lines  in the TopBase  \op\ data base
\citep{Seaton1987}, with collision strengths determined from highly approximate g-bar approximations.
Adding the \chia\ database lines into \Cloudy\ required a decision about how to integrate these
databases since some  \op\ lines may exist within \chia.
The emission spectra for the iron ions we include 
are shown in Figure \ref{fig:3spectra} and in the online material.
These show that the \op\ lines tend to occur at 
wavelengths shorter than the \chia\ lines.
\chia\ only includes lines which have collision strengths computed 
with sophisticated methods, and only for transitions with the lower level in the ground term 
(we use the g-bar approximation for subordinate lines).
The \op\ line data often extend to higher excitation levels.
We use the \chia\ data for all lines it includes, and supplement
these with higher excitation \op\ data using the g-bar approximation.
We refer to this as the Hybrid scheme.

Figure \ref{fig:3spectra} illustrates how the \op\ and \chia\ data are blended 
to give the Hybrid spectrum for one ion stage (the online material shows
all ions).
Although all calculations are done with \Cloudy, we use different
parts of the atomic database to compare spectra.
The top panel, labeled C10, shows the combination of the internal
data and the Opacity Project data that were part of C10, 
the last major release of \Cloudy.
The \chia\ spectrum (middle panel) shows only lines included in that database,
and contains more lines than \op\ (C10) at 1000\A\ and longer. 
The C10 spectrum, largely lines in the \op, 
has quite a few lines between 100\A\ and 1000\A\ that are missing from the \chia\ spectrum.
The Hybrid spectrum (lower panel) is the blending of these two spectra as described previously, 
containing both the \chia\ lines and the \op\ lines.
The Hybrid configuration will be the default in the next major release of \Cloudy.

The addition of the \op\ data to \chia\ has little effect on the cooling
in the photoionization case where
kT is relatively lower than the ionization potential.  It does increase the cooling in the collisional
case where kT approaches the ionization potential and even Rydberg levels can be excited.

\subsubsection{The \kz\ database}
\label{addkz}
\Cloudy\ uses a large model of the \feii\ emission \citep{Verner1999} and 
the H-like and He-like ions are treated in the iso-electronic approach described previously.
\feiv\ through \fexxiv\ come from \chia\ as described in \ref{cloudyhybrid}.
The only missing iron ion is Fe$^{2+}$.
To fill that void, we added data from the \kz\ database
\citep{Kurucz.R95Atomic-line-list} for that ion.
This gives energy levels and transition probabilities,
but does not contain collision strengths.
Collision strengths for the lowest 14 levels of \feiii\ are given by \citet{Zhang1996}.
For the many higher levels we use the g-bar approximation as we did for the \op.
We combine the \citet{Zhang1996} and \kz\ data with the \op\ data using the Hybrid scheme
described above.

\subsection{Iron Cooling}
\label{fecoolplots}
\begin{figure}
\includegraphics[width=\columnwidth, keepaspectratio]{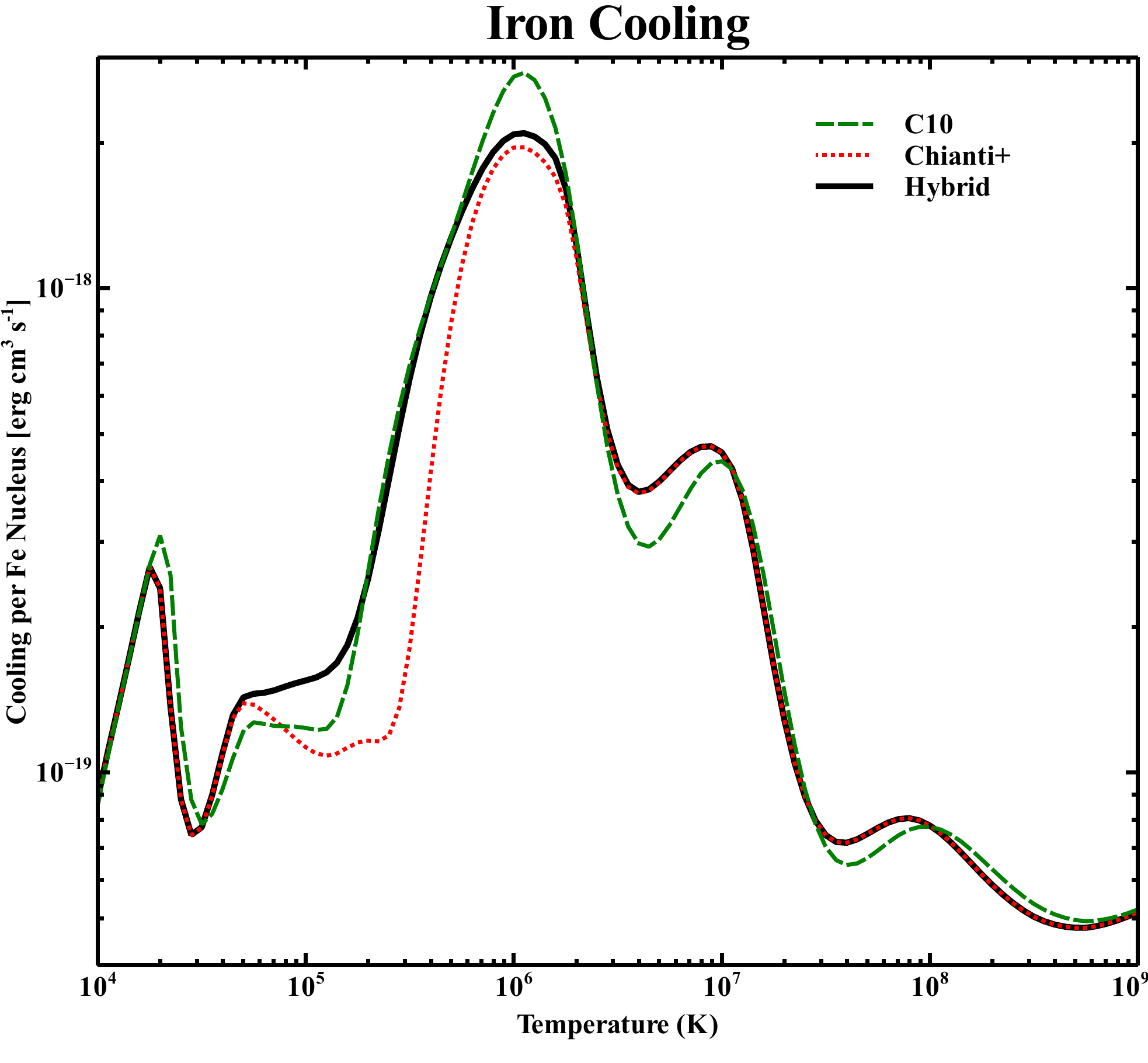}
\caption{Comparison of the iron cooling per nucleus between the ``C10'' 
internal database (the green dashed line), 
``\chia +" (\chia\ + \kz, the red dotted line), 
and ``Hybrid" (\op\ + \chia\ + \kz, the solid black line). }
\label{fig:fecoolingcloudy}
\end{figure}

As a test we computed a collisional ionization cooling curve for a
pure iron plasma using the three 
different available databases.
The results are shown in Figure \ref{fig:fecoolingcloudy}.
The green dashed line is the latest release of \Cloudy, 
known as C10. 
This uses only our internal database and the Opacity Prtoject
and does not contain any data from \chia\ or \kz. 
The red dotted line, labeled \chia +, uses \chia\ and \kz\ data, but not the \op .
The  Hybrid configuration contains \chia, the \op, and our \kz\ additions.

All three configurations have good agreement at the temperature extremes.
Between 4\e4 and 2\e5 \K, the Hybrid configuration has more cooling than the other two.
Hybrid has more cooling than C10 in this range because of the addition of the 
\feiii\ \kz\ data as well as \feiv\ and \fev\ data from \chia.
Hybrid cooling is greater than \chia+ for this temperature range and up to 5\e5 \K\ 
because it has the additional \op\ lines.

The C10 cooling exceeds both Hybrid and \chia + around 1\e6 \K. 
This is because C10 used \op\ data, with their uncertain g-bar approximation, for many 
high-excitation iron lines.
\chia\ uses real calculations of collision strengths and the values 
for the strongest lines were systematically lower
than the g-bar estimates, resulting in less cooling.
When a particular transition appears in both data sets, 
the \op\ version is used for C10 and the \chia\ version is used for both Hybrid and \chia +. 
This is why Hybrid and \chia + show equal cooling at many temperatures. 

The Hybrid and \chia + cooling are equal for temperatures greater than 1\e6 \K. 
C10 has less cooling in this range because of the additional lines in \chia. 
The C10 cooling is different than the other configurations for temperatures above 3\e7 \K\ due to different ionization distributions, caused in turn by
the updates to the recombination coefficients described in Section \ref{recombinationcoefficients} that are not present in C10.

\begin{figure}
\includegraphics[width=\columnwidth, keepaspectratio]{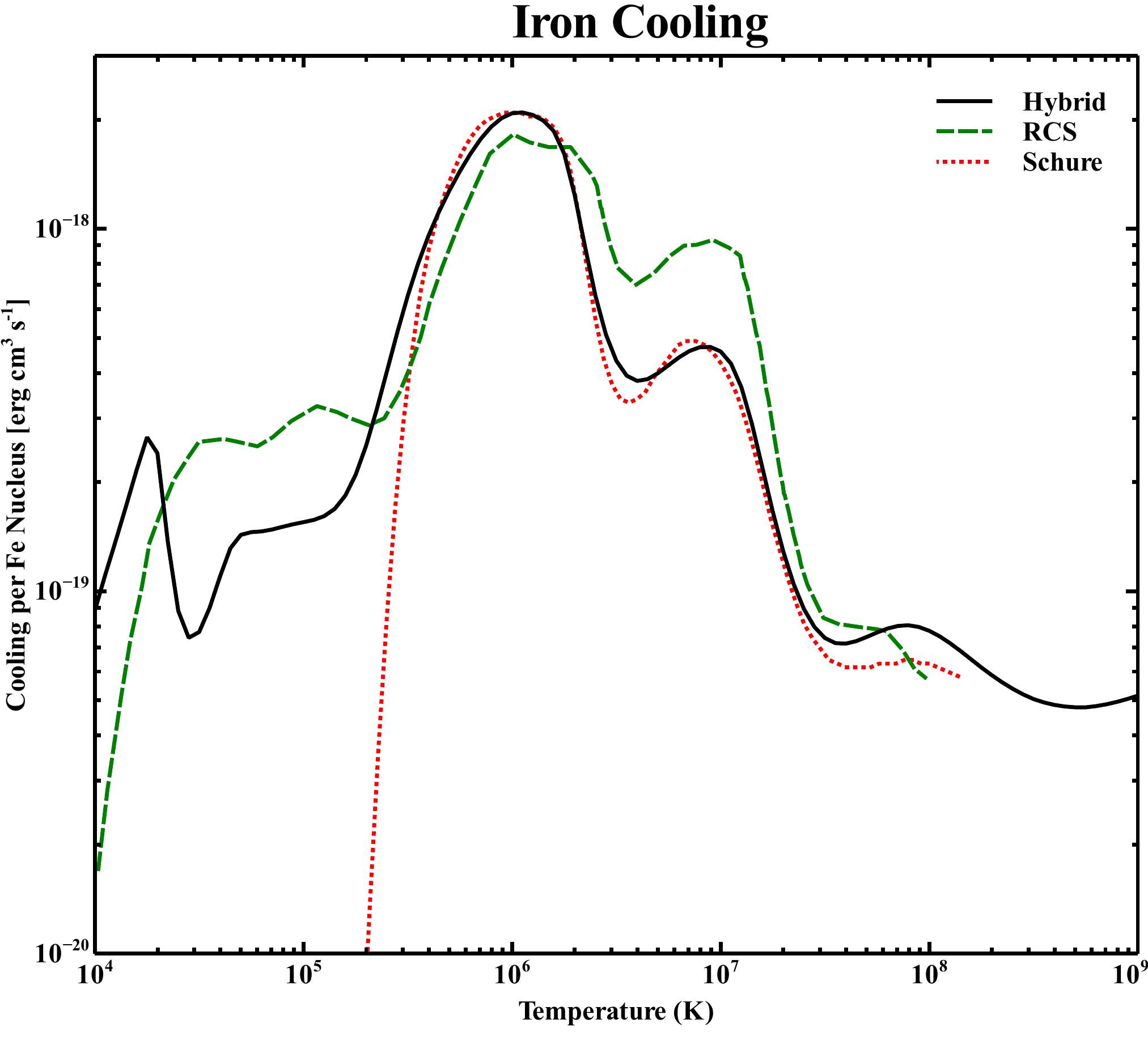}
\caption{Comparison of the iron cooling per nucleus between the Hybrid configuration of \Cloudy, \citet{Raymond1976}, and \citet{SchureEtAl09}}
\label{fig:fecoolingRCSnS}
\end{figure}

In the following sections we will compare our current calculations of the
cooling with those presented in previous works.
We concentrate on studies which report the cooling for specific elements
to remove uncertainties caused by changes in the assumed solar composition.
Figure \ref{fig:fecoolingRCSnS} compares the cooling for a 
pure iron gas for our final Hybrid configuration
with the cooling functions of \citet{Raymond1976} and \citet{SchureEtAl09}.
\citet{Raymond1976} is a standard by which many cooling functions are compared.
Our differences from \citet{Raymond1976} are likely the result of 
their use of g-bar collision strengths.
\citet{Raymond1976} used estimation techniques for all collision strengths of \fei\ through \fevii\ 
whereas we only use g-bar for \feiii, the \op\ treatment of high-excitation lines, and transitions for which \chia\ provides not collision data.
The agreement is surprisingly good considering the remarkable changes in the
atomic database in the time since their calculation.

Our Hybrid configuration is in good agreement with the \citet{SchureEtAl09} curve 
at temperatures above 3\e5 \K.
At temperatures below 3\e5 \K, the \citet{SchureEtAl09} iron cooling drops off very quickly.
\citet{SchureEtAl09} used the package SPEX, which according to the SPEX line list, does not have iron lines at ionizations less than \feviii.
This would explain their lack of iron cooling below 3\e5 \K.
Calculations of the total cooling are compared next.

\subsection{Total Cooling with our three configurations}

\begin{figure}
\includegraphics[width=\columnwidth, keepaspectratio]{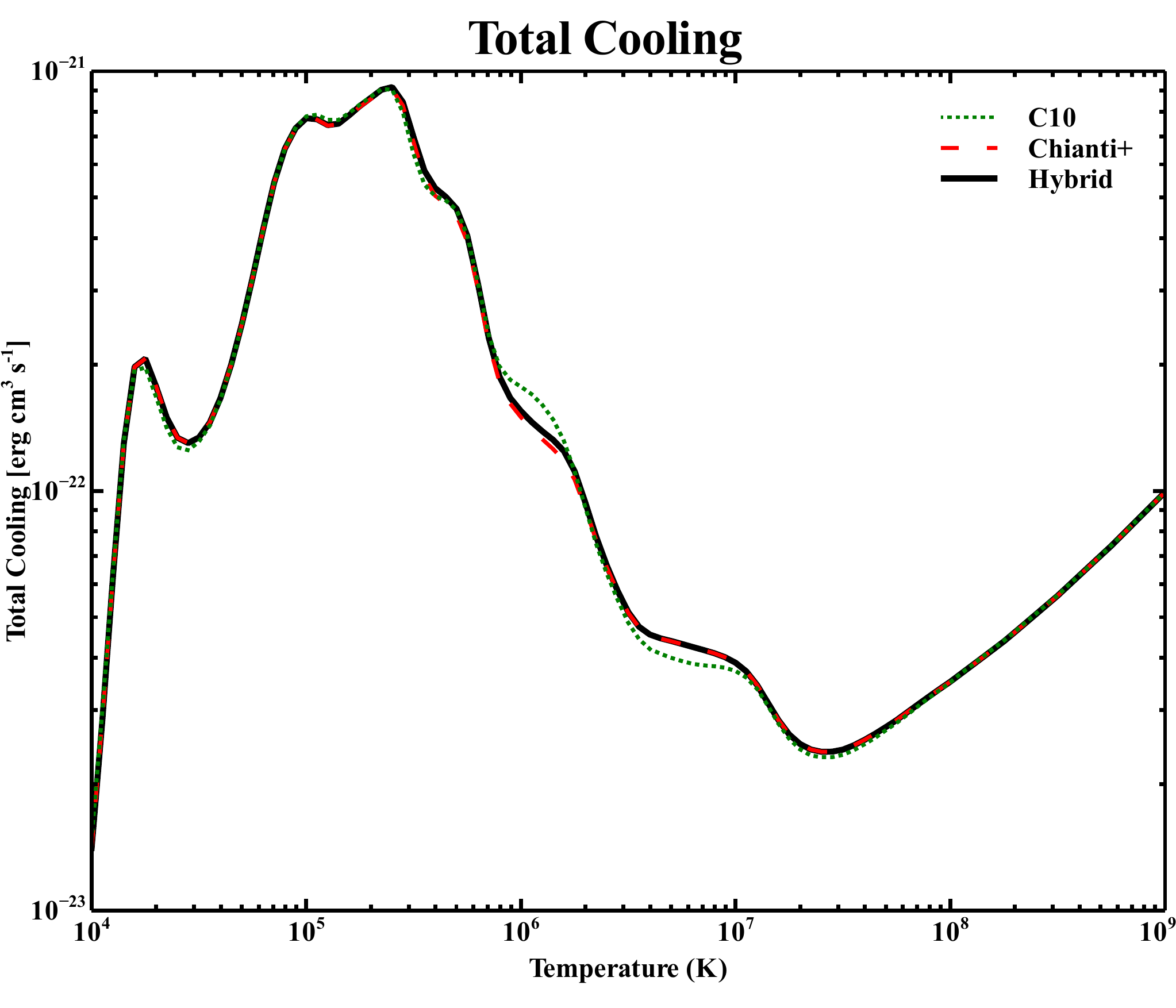}
\caption{Comparison of the total cooling between the C10, \chia+, and Hybrid configurations using abundances given in \citet{Raymond1976}}
\label{fig:totalcoolingcloudy}
\end{figure}

The previous sections focused on iron, the element we have expanded to include \chia\ data.
Here we compute the total cooling of a collisionally-ionized plasma.
This depends on all of the elements present, not just iron.
For all species other than iron we use our internal database.
Figure \ref{fig:totalcoolingcloudy} compares the total cooling for our C10, \chia+, 
and Hybrid configurations, using abundances from \citet{Raymond1976}.
(This composition was chosen to allow later comparisons with their paper.)
The \chia+ and Hybrid configurations give almost identical total cooling,
showing that our internal database is in good agreement with \chia.
They are also in reasonable agreement with C10.
The C10 total cooling is smaller than the other configurations 
between 3\e6 and 2\e7 \K\
Notice that the updated cooling predicts a region of instability 
around 6\e6~K, while the older version predicted that this
region have small regions that would have neutral thermal stablility.

\subsubsection{Comparison with \citet{Raymond1976} and \citet{SchureEtAl09}}

\begin{figure}
\includegraphics[width=\columnwidth, keepaspectratio]{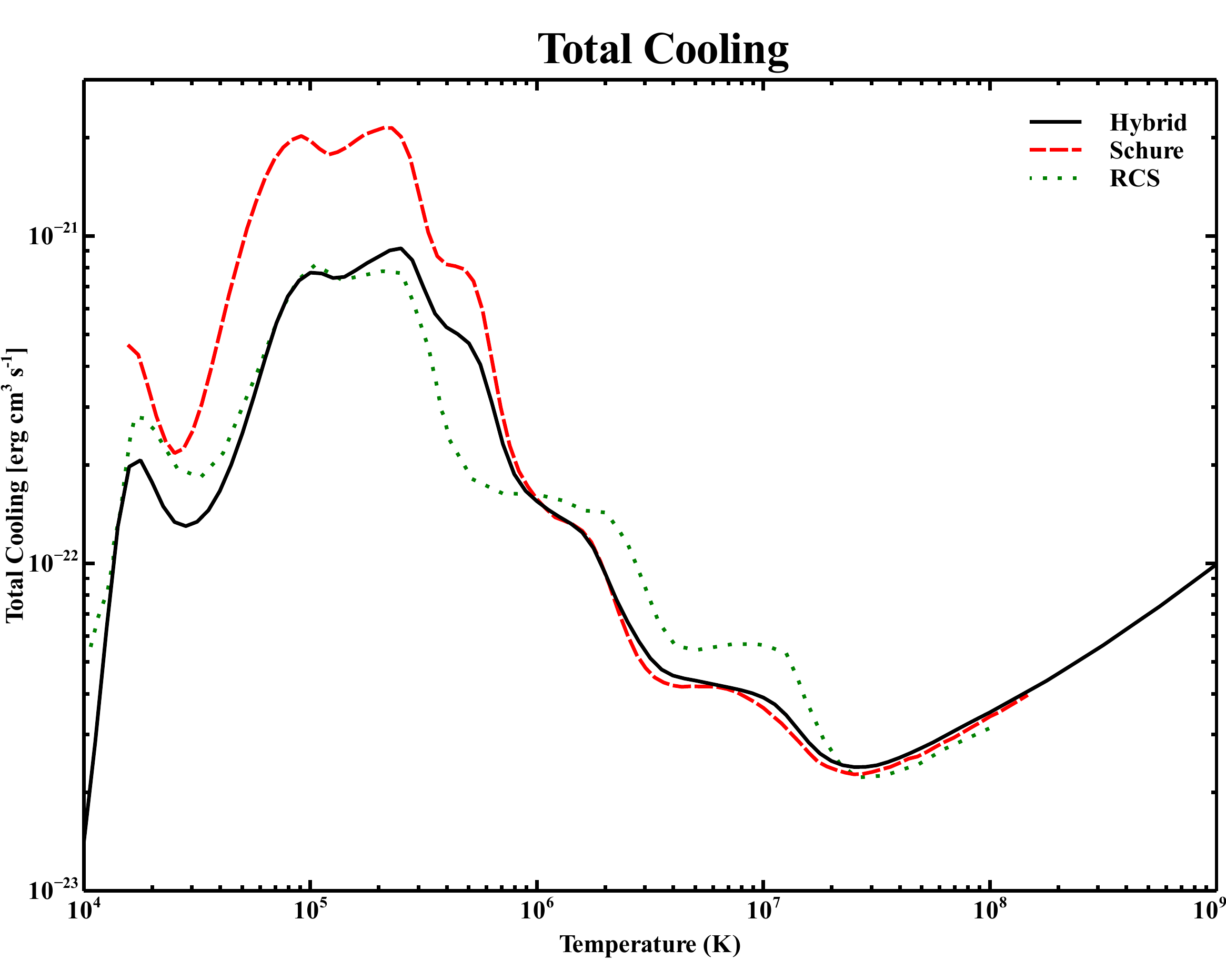}
\caption{Comparison of the total cooling between the Hybrid configuration of \Cloudy, \citet{Raymond1976}, and \citet{SchureEtAl09} using abundances of \citet{Raymond1976}}
\label{fig:totalcoolingRCSnS}
\end{figure}

We compared the iron cooling with \citet{Raymond1976} in section \ref{fecoolplots}
and Figure \ref{fig:fecoolingRCSnS}, where we found that they
produced more cooling between 2\e4 and 2\e5 \K.
Figure \ref{fig:totalcoolingRCSnS} compares 
the total cooling with all elements included.
Their total cooling is in surprisingly good agreement with our hybrid scheme
considering the major changes in the atomic data that have occurred in the
past 35 years.

Figure \ref{fig:totalcoolingRCSnS} also shows the total cooling computed by \citet{SchureEtAl09}.
We compared the iron cooling from \citet{SchureEtAl09} in section \ref{fecoolplots},
and found reasonable agreement with our Hybrid configuration 
for higher temperatures but that their
model lacked important coolants below 2\e5 \K.
The significantly higher total cooling of \citet{SchureEtAl09} around 10$^5$ K
cannot be due to iron.
Section \ref{schure} explains some possible reasons for this difference.

\subsubsection{Comparison with \citet{Sutherland1993} and \citet{Foster2012}}

\begin{figure}
\includegraphics[width=\columnwidth, keepaspectratio]{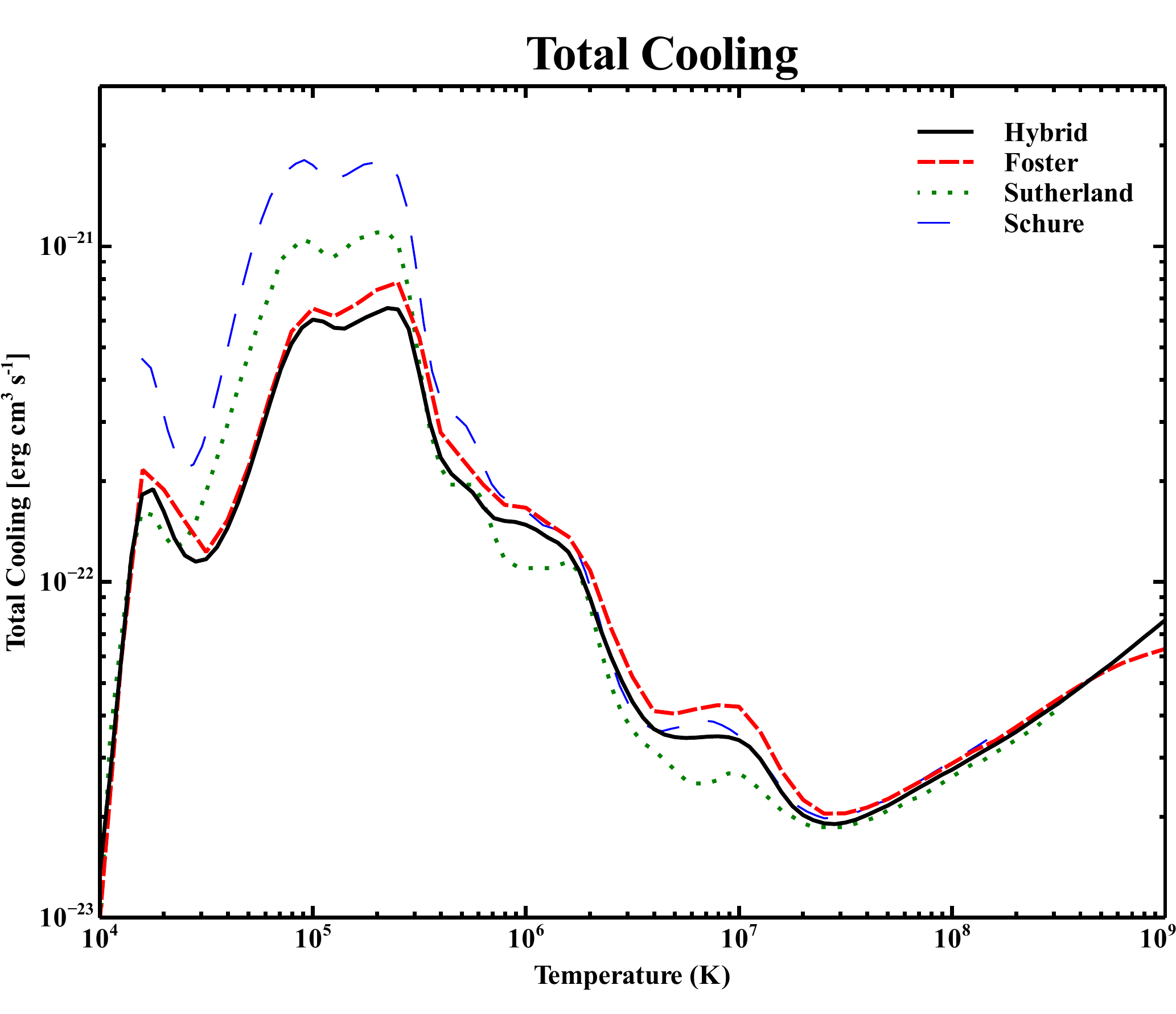}
\caption{Comparison of the total cooling between our Hybrid configuration, \citet{Foster2012},
\citet{Sutherland1993}, and \citet{SchureEtAl09} using abundances of \citet{Anders1989}}
\label{fig:totalcoolingSutherland}
\end{figure}

\citet{Foster2012} describes the latest additions to AtomDB, 
an atomic database that focuses on X-ray astronomy,
and which, like \chia\ and \Cloudy, pays particular attention to the atomic physics. 
In addition to describing all of the atomic data updated in the latest release of AtomDB, \citet{Foster2012} also provides a total cooling function based on solar abundances of \citet{Anders1989}.
\citet{Sutherland1993} used MAPPINGS II to produce cooling functions between 1\e4 and 1\e{8.5} \K.
MAPPINGS II includes calculations for 16 elements with all ion stages.

Figure \ref{fig:totalcoolingSutherland} compares our Hybrid total cooling, \citet{Foster2012},
\citet{SchureEtAl09}, and \citet{Sutherland1993}, 
using the common solar abundances  of \citet{Anders1989}.
The four cooling functions have a similar overall shape.
We agree very well with \citet{Foster2012} at all temperatures, specifically around 1\e5 \K\ where we differ significantly from \citet{SchureEtAl09}.
However, the cooling at the peak near 1\e5 \K\ ranges from Hybrid to about a factor of two larger.
The Sutherland cooling function lies roughly midway between our Hybrid and
\citet{SchureEtAl09}.
The element or elements causing the difference 
around 2\e5 \K\ between \Cloudy\ and \citet{SchureEtAl09} are possibly the same reason for the difference with \citet{Sutherland1993}.
The differences with \citet{SchureEtAl09} are more extreme and we concentrate on that in Section \ref{schure}.

\subsubsection{Comparison with \citet{Colgan2008}}

\begin{figure}
\includegraphics[width=\columnwidth, keepaspectratio]{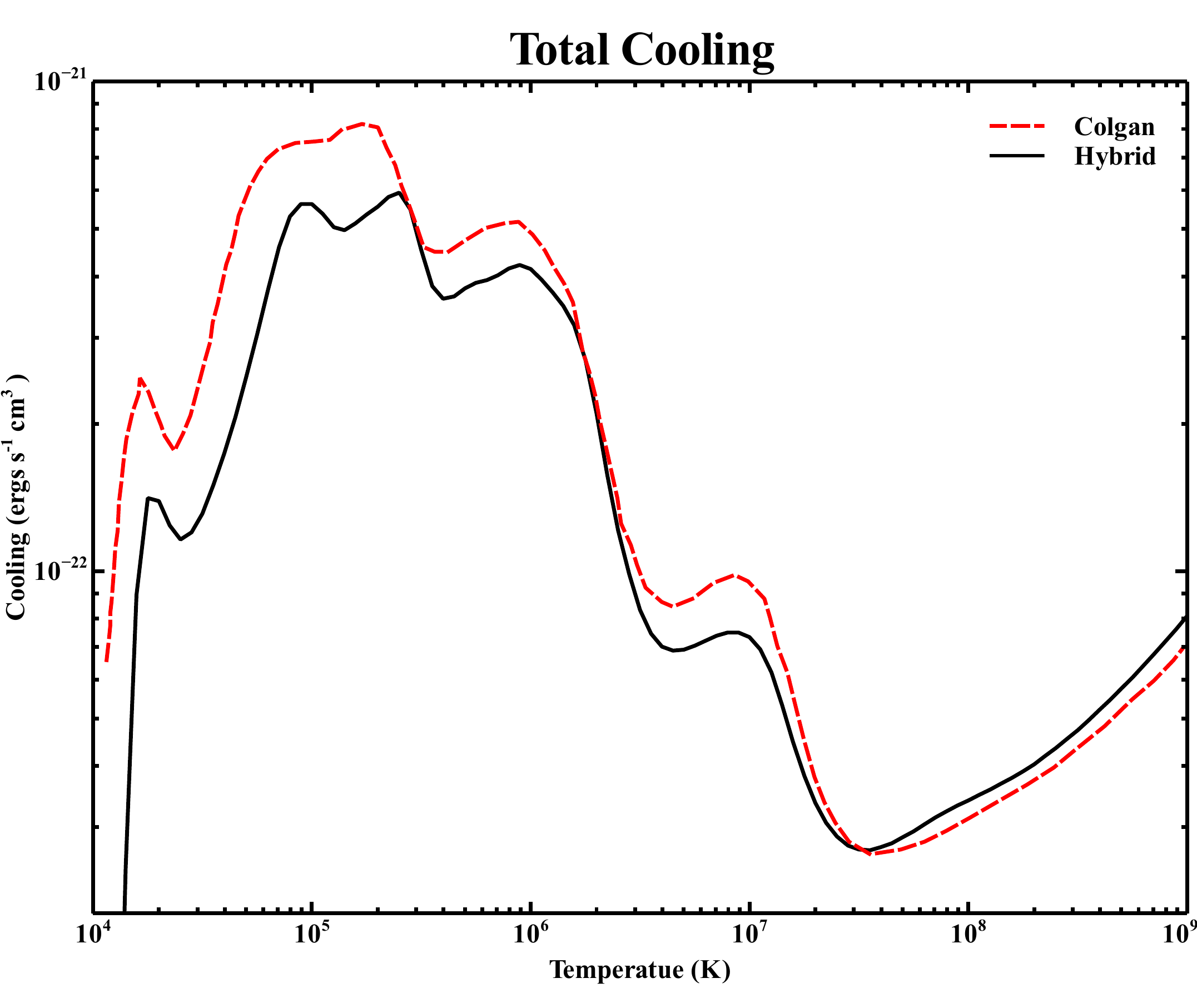}
\caption{Comparison of the total cooling between Hybrid and \citet{Colgan2008} using the abundances of \citet{Colgan2008}}
\label{fig:totalcoolingColgan}
\end{figure}

\citet{Colgan2008} used the Los Alamos plasma kinetics code ATOMIC to calculate radiative losses for a specific set of abundances. 
They also used several programs that are part of the Los Alamos suite of atomic structure and collision codes to generate the data needed to calculate the losses. 
Figure \ref{fig:totalcoolingColgan} compares the Hybrid total cooling function with \citet{Colgan2008}.  
Note that the \citet{Colgan2008} plot has been converted from Watts to ergs.
\citet{Colgan2008} find significantly more cooling around 1\e5 \K.
\citet{Colgan2008} also compared their total radiative losses with a similar calculation
using data from \chia\ version 6.
Their \chia\ 6 results are in reasonable agreement with our Hybrid cooling.

\subsubsection{Differences with \citet{SchureEtAl09}}
\label{schure}
The \citet{SchureEtAl09} cooling curve was produced using the SPEX package
using solar abundances from \citet{Anders1989}.
They provide cooling rates for each element so that their results can be 
scaled to fit any set of abundances.
We used these individual cooling rates to include the \citet{SchureEtAl09} results 
to Figures \ref{fig:totalcoolingRCSnS}, \ref{fig:totalcoolingSutherland}, and \ref{fig:totalcoolingBest}.
Their calculation shows greater cooling than our Hybrid configuration 
for $T < 10^6\K$.

We looked into individual coolants to find the reason for this difference.
It is not due to iron since the \citet{SchureEtAl09} iron cooling significantly less than Hybrid around 1\e5 \K\  in Figure \ref{fig:fecoolingRCSnS}.
\Cloudy\ reports that the dominant coolants around 1\e5 \K\ are carbon and oxygen.
In the next section we examine these coolants more closely.

\subsection{Carbon and Oxygen Cooling}

\begin{figure}
\includegraphics[width=\columnwidth, keepaspectratio]{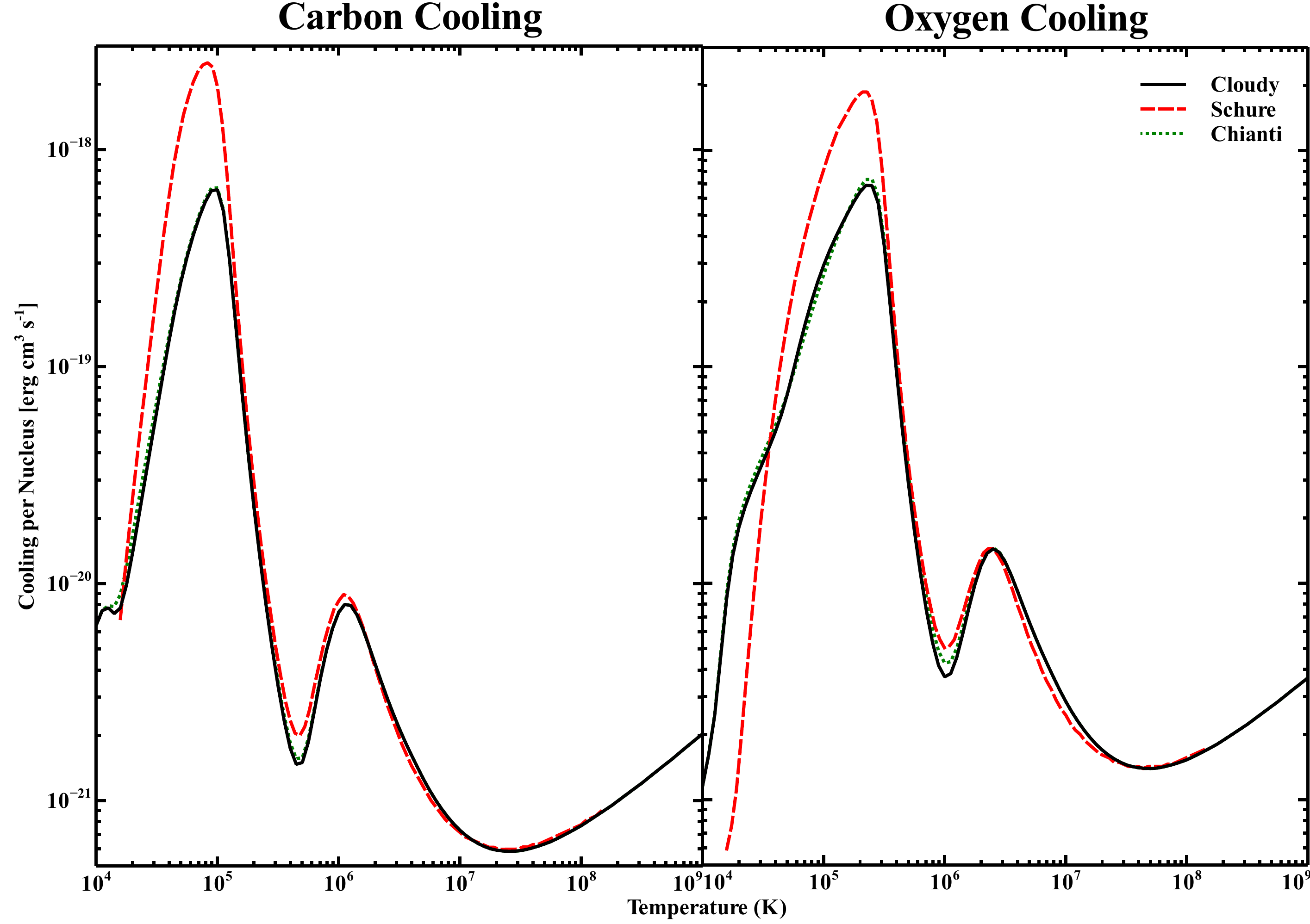}
\caption{Comparison of the carbon cooling per nucleus (left) and the oxygen cooling per nucleus (right) for Hybrid, \citet{SchureEtAl09}, and \chia.}
\label{fig:carbonoxygencooling}
\end{figure}

The comparison presented above shows that the largest discrepancies 
occur around 1\e5 \K, regions where the dominant coolants 
are carbon and oxygen.
Figure \ref{fig:carbonoxygencooling} compares the carbon and oxygen cooling per nucleus for our Hybrid, \citet{SchureEtAl09},  and \chia\ version 7.
Our Hybrid model, which only uses our internal database for these elements,
is in good agreement with \chia.
These show that \citet{SchureEtAl09} predict significantly higher cooling
below $10^6\K$, which accounts for the differences in the total cooling.

The collision strengths are most likely to be the source of the differences in 
the cooling.
We find that the primary carbon cooling transitions at 1\e5 \K\ are 
977\A\ of C III and 1548\A\ of C IV.
For these transitions, \Cloudy\ uses collision strengths from 
\citet{Berrington1985} and \citet{Cochrane1983} respectively.
The C III 977\A\ transition is the dominant coolant, 
contributing 56\%\ of the carbon cooling,
while the C IV 1548\A\ line contributes about 20\%.

The oxygen cooling at 2\e5 \K\ is dominated by the 630\A\ line of O V and a multiplet 
of 4 O IV lines around 554\A.
\Cloudy\ uses collision strengths from \citet{Berrington1985} for 630\A.
The 554\A\ transitions come from the \op\ which means that the collision strengths 
are generated using the g-bar approximation.
The dominant line is O V 630\A\ with 40\%\ of the oxygen cooling while the 
O IV 554\A\ lines account for 17\% of the total.

It is clear from Figure \ref{fig:carbonoxygencooling} that the carbon and oxygen cooling for \Cloudy\ and \chia\ are very similar over the entire temperature range,
despite being completely independent implementations of the atomic physics. 
We compared the sources for the collision data as a check on its reliability. 
For the 977 and 1548 \A\ carbon transitions, \chia\ uses collision strengths from \citet{Berrington1985} and \citet{Griffin2000} respectively.
\Cloudy\ and \chia\ use the same source for the 977 \A\ transition.
The \Cloudy\ and \chia\ values for the C IV 1548\A\ collision strength differ by only 6\%.

The results are similar for the oxygen transitions.
The \chia\ collision strength for 630\A\ is 3\%\ less than \Cloudy\ and comes from Berrington (2003; private communication).
The 554\A\ multiplet uses g-bar collision strengths in \Cloudy\
while \chia\ use values from \citet{Zhang1994}.
The values of the collision strengths differ by only about 5\%.

The fact that \citet{SchureEtAl09} find more cooling than 
\Cloudy\ or \chia\ could be explained if they included important cooling 
lines which are not present in \Cloudy\ or \chia.
C$^{2+}$ and O$^{3+}$ are the dominant stages of ionization at 
1\e5\K\ and 2\e5\K\ respectively.
The collisional ionization distribution for hydrogen and helium can be found in Table \ref{tab:iondist} and for the lightest 30 elements the online version of the table.
\citet{SchureEtAl09} used SPEX for its calculations and we were able to compare the SPEX line list to ours and with \chia.
While \chia\ has significantly more transitions, there are several transitions in the SPEX line list that are not in \chia.
These transitions are in the tens or few hundreds of Angstroms in wavelength.
At temperatures around 1\e5 \K \
these transitions cannot contribute much cooling due to their small Boltzmann factor.
This comparison suggests that the collision rates for the lines mentioned above
is the source of the differences.

Unfortunately we cannot compare our collision data with those of \citet{SchureEtAl09}, 
\citet{Sutherland1993},  and \citet{Colgan2008} 
because they do not cite sources for,
nor give values of, their atomic data.
They provide plots for element cooling but not a table of values. 
The best we can do is to say that three of the four dominant cooling lines around 1\e5 and 2\e5 \K\ have two independent sources for collision strengths and that they 
are in good agreement.
This provides us with confidence in our atomic data as well as our cooling functions.

The iron cooling is in good agreement with the cited references above.
This is the dominant coolant for higher temperatures.
Differences in the carbon and oxygen cooling account for the differences we notice
at lower temperatures.
The last step is to compare the cooling predicted using
\Cloudy's internal atomic data set to \chia.

\subsection{Comparison with the \chia\ 7 cooling function}

In the hybrid implementation we use \chia\ for iron but our internal
database for other elements.
Here we compare the cooling computed using our database with \chia.
Since we are using \chia's iron data in \Cloudy\ Hybrid, 
we exclude iron in this comparison.

\begin{figure}
\includegraphics[width=\columnwidth, keepaspectratio]{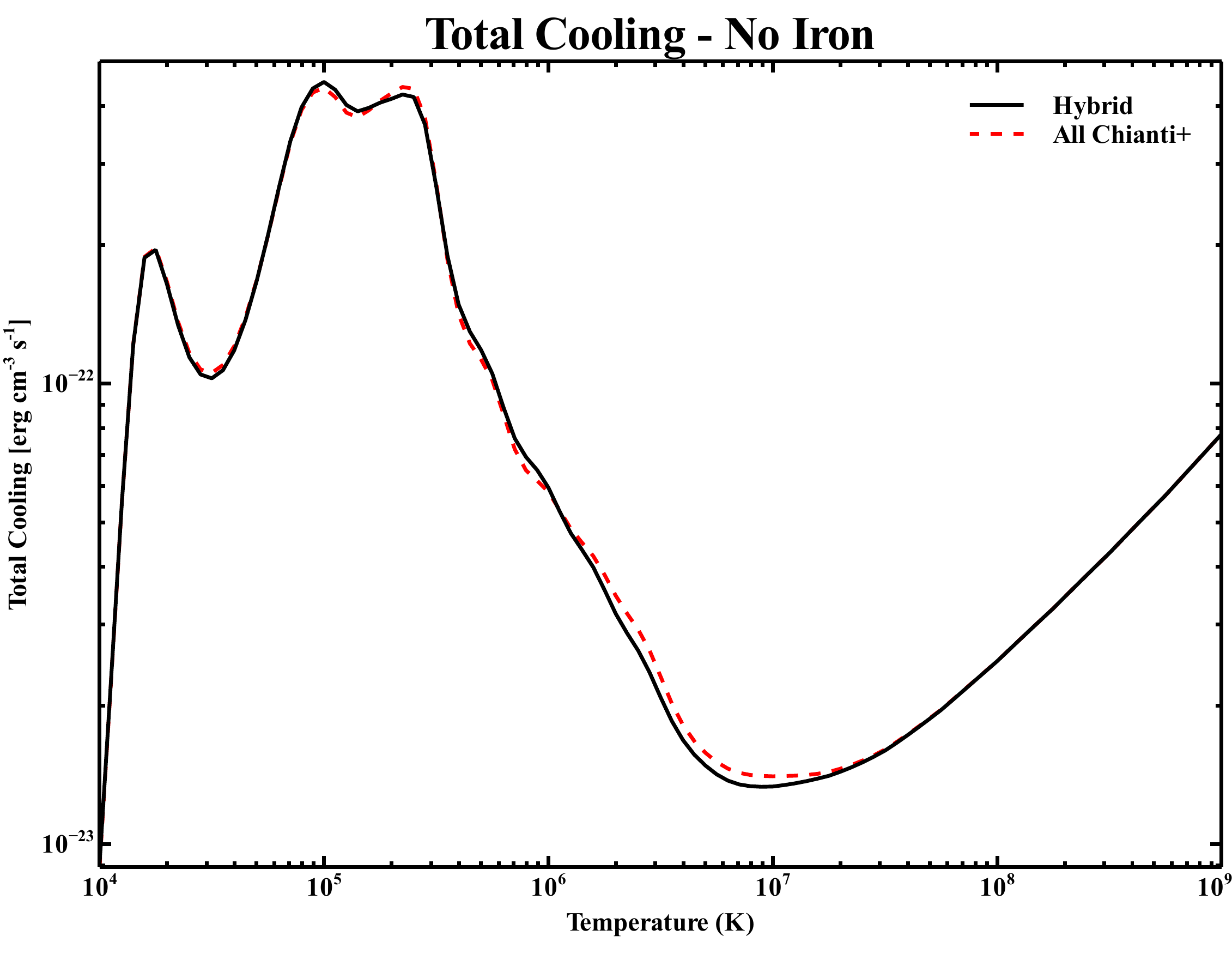}
\caption{Comparison of the total cooling with the 
Cloudy internal database and most of \chia\ version 7, labelled  "All Chianti+". 
\citet{Grevesse2010} abundances are used except that iron is not included.
Other than the H and He-like iso-sequences, where our treatment is used for both,
these curves use fully
independent implementations of the current atomic physics literature.
The good agreement suggests that these implementations are complete
and in accordance with one another.}
\label{fig:totalcoolingChianti}
\end{figure}

Figure \ref{fig:totalcoolingChianti} shows this comparison.
The \chia\  calculation uses their data for all other species 
other than the H and He-like iso-sequences where our internal models are used.
Since these two calculations use independent implementations of the
atomic database,
the good agreement between the
cooling for all temperatures shows that 
these databases are in good agreement.

\subsection{Total cooling in the collisional and photo ionization cases}

\begin{figure}
\includegraphics[width=\columnwidth, keepaspectratio]{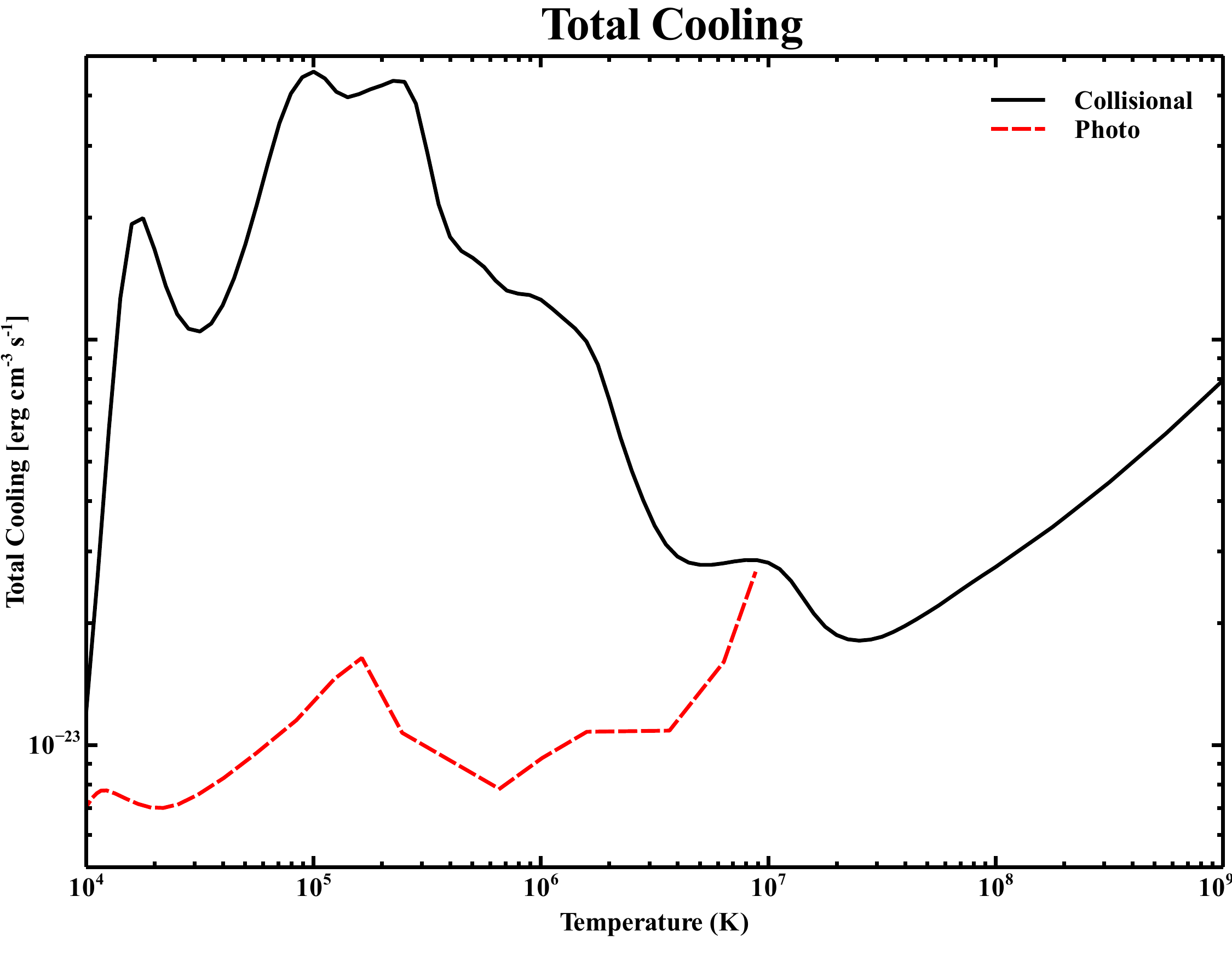}
\caption{The total cooling for the collisional and photo cases of Hybrid \Cloudy\ using solar abundances of \citet{Grevesse2010}}
\label{fig:totalcoolingBest}
\end{figure}

The goal of this paper is to establish the cooling function used in the
Williams et al.\ (in preparation) calculation of the spectrum of a cooling
non-equilibrium plasma. 
Figure \ref{fig:totalcoolingBest} shows our best estimate of the total cooling
for the collisional and photoionization cases, using solar abundances from \citet{Grevesse2010}.
Appendix \ref{inputscripts} provides the \Cloudy\ commands used to create these plots as well as a description of each command.
The cooling function for the collisional case is essentially the same as the as the Hybrid cooling functions we have shown in previous plots but with 
more recent solar abundances.
Notice that, with this mix of abundance, the region around 5\e6 K is
thermally stable, whereas it was marginally stable with previous
abundance sets.
The collisional cooling is given on a per element basis in the Appendix \ref{coolingtable} and a tab-delimited version is available in the online version. 
\citet{Gnat2012} provides element by element cooling using the C10 version of \Cloudy.
The differences between cooling plots in \citet{Gnat2012} and C10 in Figure \ref{fig:fecoolingcloudy} are due to using different abundances.

The discussion of Figure \ref{fig:tmaxvsion} explains how 
the photoionization cooling function was calculated.
We used the same SED and varied the log of the ionization parameter
between -6 and 3, which is the same as Figure \ref{fig:tmaxvsion}.
The minimum temperature in the plot is 1\e4 \K, 
to be similar to previous Figures,
although the kinetic temperature goes to lower values for low ionization parameter.

The cooling in the photoionization case is lower than the collisional case
at each temperature.
The gas kinetic temperature does not play a fundamental a role in
photoionization equilibrium, because the ionization is determined by the
balance between photoionization and recombination, which have only weak
temperature dependences.
The cooling rate in the photoionization case is determined by the heating rate,
which is set by the SED (AGN3 Chapter 3; \cite{Ferland2003b}). 
The temperature is the result of the interplay between this heating and the
gas composition, the so-called ``thermostat effect'' in photoionization equilibrium
(AGN3).
The result is the significantly lower kinetic temperature, as shown
in both Figures \ref{fig:tmaxvsion} and \ref{fig:totalcoolingBest}.
This distinction will play a major role in our selection of the the default number
of levels used to compute the spectra in our new level trimming feature, described below.

\subsection{Level trimming}
\label{levellimits}

In addition to the experimental iron data, \Cloudy\ can also use the theoretical iron data provided by \chia.
All of the tables and figures in this paper, with the exception of this section, use only experimental \chia\ data with no trimming.
This section is provided to describe the new capabilities of \Cloudy\ and does not affect any other section of the paper.
Some \chia\ ions which are of particular interest in the solar case
have  more than 300 theoretical energy levels and some have as many as 700 levels.
The \kz\ database often has thousands of levels.
Solving the level populations is quite expensive due to the large number
of evaluations of the emission and cooling during the solution of the
equations of statistical, thermal, and ionization equilibrium.

We added an option to restrict the maximum number of energy levels 
that are used for each species.
There is a tradeoff between including more levels and lines, 
which provides a more accurate simulation but with longer run times,
versus more compact models, with fewer levels and lines and shorter
compute times, but perhaps with some loss of fidelity. 
This section outlines how we chose the default number of levels,
and shows the effects this has on predicted quantities.
It is important to note that the level limiting only applies to the newly 
added \chia\ and \kz\ data.
All of the data used in C10, including \op\ data, are unaffected by this limiting.

\begin{figure}
\includegraphics[width=\columnwidth, keepaspectratio]{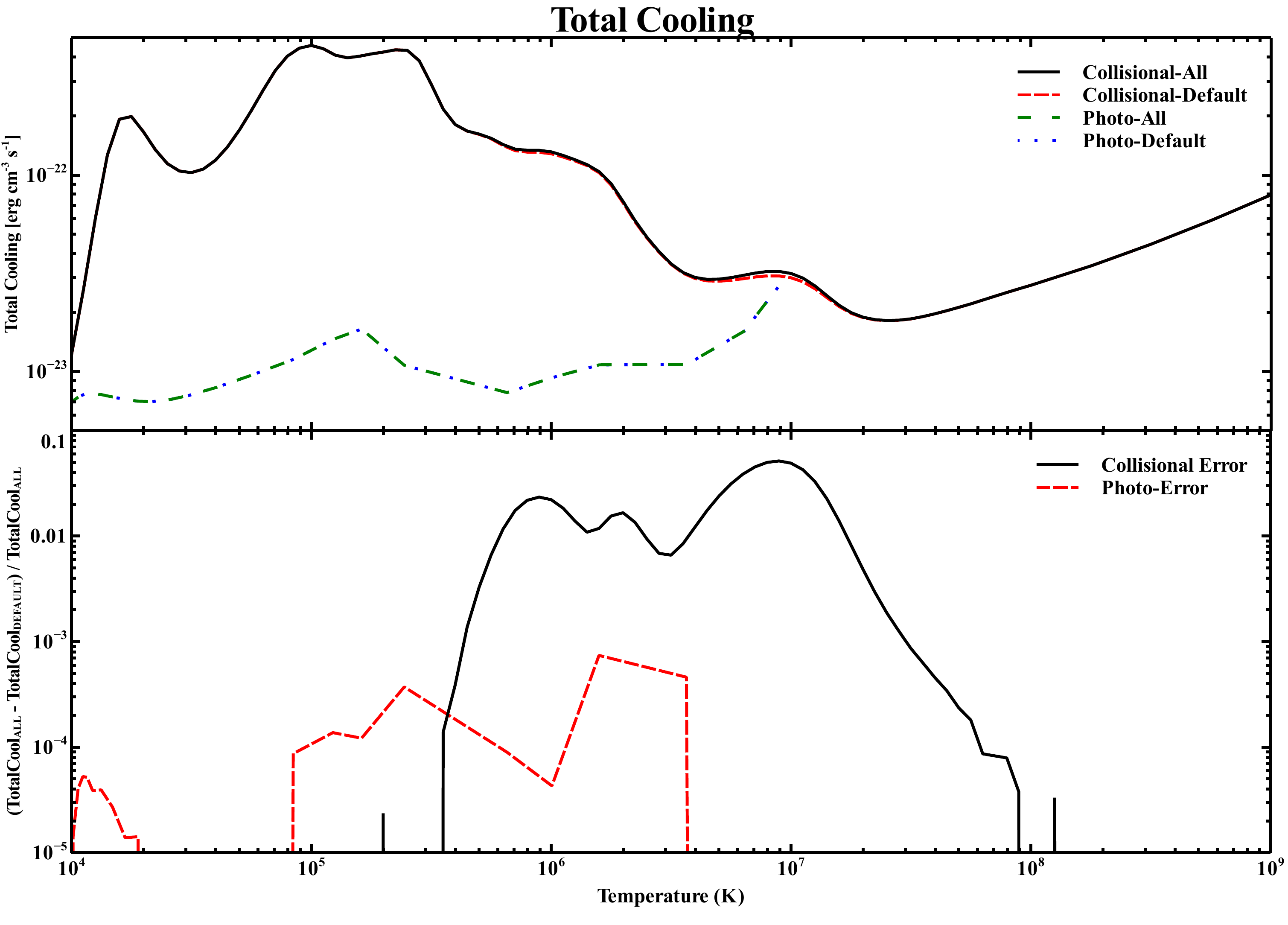}
\caption{Total cooling using \chia\ theoretical energy levels of \Cloudy\ Hybrid collision and photon dominated cases using all available levels and the default numbers (top) and the associated 
relative difference  (bottom) described in Section \ref{levellimits}}
\label{fig:LevelLimitComb}
\end{figure}

Figure \ref{fig:LevelLimitComb} shows the total cooling $\Lambda_{ALL}$
(top panel) using
all theoretical levels in the \chia\ and \kz\ databases, and the cooling with a particular
subset of these levels described below.
The differences are small.
The lower panel shows the relative difference between
the full database  and our compact model with $n$ levels as
$(\Lambda_{ALL} - \Lambda_{n} ) /  \Lambda_{ALL}$.
Plots like this one were used to decide on an optimum default limit 
to the number of levels $n$.

The kinetic temperature in a photoionized gas is lower, for a particular
ionization state, than in a collisional gas (Figure \ref{fig:tmaxvsion}).
As a result, for a given ion, more levels will be energetically accessible in
the collisional case compared with the photoionized case.
Since we will be adding more of the \chia\ species in the future and since \chia\ iron species have significantly more energy levels
than most other species, we have a default number of levels for iron species and one for all other species.
After some experiments we settled on a default limit of
100 levels for the collisional case 
and 25 levels for the photoionization case for each iron species.
For all other \chia\ species, which will be added in the future, the default limits are 50 for collisional and 15 for photionization.
We find that these limits capture nearly all of the total cooling 
but requires about five times less compute time than using the full database.
This is the approximation shown in
the top panel of Figure \ref{fig:LevelLimitComb}.
For the collisional case the limited levels produces cooling within
1\% of the total for most temperatures, with the worst agreement 
of $\sim 5\%$ around
${10^7}$ K.
This peak error is mostly due to limiting \fexix\ to 100 of its 636 levels. 
The photoionization cooling functions are essentially identical with the smaller
number of levels reproducing the total cooling within better than 0.1\%.

Since the accuracy required for a specific simulation may depend on particular goals,
we provide a simple input option to change the number of levels.
All of the plots in this paper are using the full \chia\ and \kz\ databases 
unless otherwise noted.

\section{Conclusions}

This paper outlines some improvements to the plasma simulation code
\Cloudy.
It will become the reference for future improvements in the atomic and
molecular database.
The specific results are the following:

\begin{itemize}

\item
We added much of the experimental iron data provided by \chia\ version 7 \citep{Landi2012}.
In addition, we  added energy levels and transition probabilities from the 
\kz\ database \citep{Kurucz2009} for \feiii.
Those \feiii\ data were supplemented by collision strengths using the g-bar approximation.
\feii\ is treated with the model described by \citep{Verner1999} and
the H-like and He-like ions are treated with the unified described
by \citet{PorterFerland2007}.

\item
The Hybrid configuration was added to expand the wavelength coverage of \Cloudy\ by merging  the existing \op\ data, which we have long included, with the  \chia\ and \kz\ data.
The \op\ and \kz\ data do not have corresponding collision rates, and we have
used the g-bar approximation to include them.

\item
We  updated \Cloudy\ with the latest recombination coefficients from 
the Badnell website, including \citet{Badnell2010} and
\citet{Abdel-Naby.S12Dielectronic-recombination-data-for-dynamic}.
We updated \Cloudy 's collisional ionization rate coefficients to those 
given in \cite{Dere.K07Ionization-rate-coefficients-for-the-elements},
although these are very similar to the rates of \citet{Voronov1997} which we have
used since soon after the publication of that paper.
The ionization distributions shown in this paper are based on these updates.

\item
We added an option to limit the number of energy levels that will 
be used for a particular simulation in order to reduce run time 
without sacrificing accuracy.
This reproduces the results of the full databases to within much better than 5\% but with five times shorted run times.
It is easy to change this option with user accessible commands.
This limiting was not used to generate any figures or tables in this paper except for Figure \ref{fig:LevelLimitComb}, which
demonstrates the accuracy of the level trimming results.

\item
Our iron cooling function created with the Hybrid mix 
of the \op, \chia, and \kz\ data agrees surprisingly well with that of \citet{Raymond1976} considering the remarkable changes in
the atomic database in the past 35 years.
Significant differences exist with
\citet{SchureEtAl09} 
at temperatures below 1\e6 \K.
We attribute the difference in iron cooling to missing
ions (\fevii\ and below) in the SPEX tool.

\item
The total cooling also agrees with \citet{Foster2012} and when we use
the full \chia\ database.
We find less cooling around 1\e5 \K\ compared to \citet{SchureEtAl09}, 
\citet{Sutherland1993}, and \citet{Colgan2008}.
Detailed comparisons show that differences with \citet{SchureEtAl09} 
(the calculation where such comparisons are possible)
are due to carbon and oxygen cooling.
We compared our Hybrid configuration total cooling to the total cooling 
using the \chia\ version 7 data for the majority of the available species and concluded that these two 
implementations of the atomic data literature are in good agreement.
We suspect that the differences in cooling compared with 
\citet{SchureEtAl09}, \citet{Sutherland1993}, and \citet{Colgan2008}
are due to their atomic data, 
although it is not possible to track down what they used.

\item
\Cloudy\ is open source and is freely available.  
Appendix \ref{inputscripts} in the on-line version provides the \Cloudy\ input scripts
needed to generate the cooling functions used in this paper.

\item
We provide detailed ionization and cooling rates for each of the thirty elements
included in the calculation in the on-line version.

\end{itemize}

GJF acknowledges support by NSF (0908877; 1108928; and 1109061), 
NASA (07-ATFP07-0124, 10-ATP10-0053, and 10-ADAP10-0073), 
JPL (RSA No 1430426), and 
STScI (HST-AR-12125.01, GO-12560, and HST-GO-12309).
PvH acknowledges support from the Belgian Science Policy Office
through the ESA Prodex programme.

\clearpage

\bibliography{biblioLocal,./common/bibliography2}

\clearpage

\appendix
\section{Input Scripts}
\label{inputscripts}

We have included the \Cloudy\ input scripts for calculating the cooling in both collisional and photon dominated cases.
These are the scripts used to generate Figure \ref{fig:totalcoolingBest}, which assumes solar abundances.
We have also provided a description of each command.

\vspace{1cm}

\noindent\begin{minipage}{\textwidth}
\captionof{table}{Collision dominated cooling of \Cloudy\ Hybrid using solar abundances}
\label{tab:collscript}
\centering
\begin{tabular}{ |  l | }
\hline
coronal 4 vary\\
atom chianti hybrid "CloudyChiantiKurucz.ini"\\
abundances GASS10\\
atom feii\\
grid 4 9 0.05\\
hden 0\\
stop zone 1\\
set dr 0\\
set eden 0\\
save cooling "hybrid-coll.col" last no hash\\
\hline
\end{tabular}
\end{minipage}

\vspace{1cm}

\noindent\begin{minipage}{\textwidth}
\captionof{table}{ Photon dominated cooling of \Cloudy\ Hybrid using solar abundances}
\label{tab:photoscript}
\centering
\begin{tabular}{ |  l | }
\hline
table agn\\
atom chianti hybrid "CloudyChiantiKurucz.ini"\\
abundances GASS10\\
ionization parameter -2 vary\\
grid -6 3 0.25\\
hden 0\\
stop zone 1\\
save cooling "hybrid-photo.col" last\\
\hline
\end{tabular}
\end{minipage}

\vspace{1cm}
\clearpage

\begin{description}
\item [\bf coronal 4 vary] \hfill \\
Coronal sets up a collisonally ionized gas at 1\e4 \K\ and vary it based on the grid command.

\item [\bf atom chianti hybrid "CloudyChiantiKurucz.ini"] \hfill \\
This command enables Hybrid mode using all species listed in CloudyChiantiKurucz.ini.

\item [\bf abundances GASS10]\hfill \\
This makes \Cloudy\ use the solar abundances from \citet{Grevesse2010}.

\item [\bf atom feii]\hfill \\
Atom feii enables the \feii\ model developed by \citet{Verner1999}.

\item [\bf grid 4 9 0.05]\hfill \\
The grid command gives the limits of what is being varied as well as the increment.
For Table \ref{tab:collscript}, the temperature is being varied between 1\e4 and 1\e9 \K\ in 0.05 dex increments.
Table \ref{tab:photoscript} is varying the ionization parameter.

\item [\bf hden 0]\hfill \\
Hden sets the log of the total hydrogen density.
In this case, it is set to $1 \pcc$.

\item [\bf stop zone 1]\hfill \\
Stop zone sets the limit to the number of zone to calculate per iteration.

\item [\bf set dr 0]\hfill \\
Set dr sets the log of the zone thickness in cm.

\item [\bf set eden 0]\hfill \\
Set eden sets the log of the electron density in $ \pcc $.

\item [\bf table agn]\hfill \\
Table AGN sets up the incident radiation field using a continuum from \citet{MathewsFerland87}.

\item [\bf ionization parameter -2 vary]\hfill \\
The ionization parameter is the dimensionless ratio of hydrogen-ionizing photon to total-hydrogen
densities.

\item [\bf save cooling "hybrid-coll.col" last]\hfill \\
This saves the cooling agents for the last zone.
\end{description}

\clearpage

\begin{table} \section{Cooling by Element}\label{coolingtable}
  \centering
  \caption{Element specific cooling [erg cm$^3$ s$^{-1}$ for the lightest 30 elements which come from revision 6417 of the \Cloudy\ trunk.}
    \begin{tabular}{ccccccccccc}
\hline
\hline
    Te (K)  & Hydrogen & Helium & Lithium & Beryllium & Boron & Carbon & Nitrogen & Oxygen & Fluorine & Neon \\
\hline
1.00E+04 & 4.32E-24 & 6.85E-29 & 6.80E-22 & 4.93E-21 & 3.16E-22 & 6.41E-21 & 3.11E-21 & 1.15E-21 & 2.53E-25 & 7.63E-28 \\
1.58E+04 & 1.71E-22 & 1.26E-25 & 1.01E-22 & 2.14E-19 & 1.58E-20 & 7.56E-21 & 1.37E-20 & 9.04E-21 & 1.17E-21 & 1.01E-23 \\
2.51E+04 & 7.93E-23 & 2.48E-23 & 2.67E-23 & 2.09E-19 & 1.61E-19 & 2.93E-20 & 3.37E-20 & 2.66E-20 & 1.06E-20 & 6.10E-22 \\
3.98E+04 & 2.82E-23 & 5.02E-23 & 1.04E-23 & 1.20E-20 & 6.66E-19 & 1.30E-19 & 8.05E-20 & 4.98E-20 & 1.76E-20 & 1.34E-21 \\
6.31E+04 & 1.28E-23 & 3.58E-22 & 2.14E-23 & 1.30E-21 & 4.77E-19 & 4.01E-19 & 1.98E-19 & 1.25E-19 & 5.54E-20 & 2.00E-20 \\
1.00E+05 & 6.98E-24 & 6.21E-22 & 5.59E-22 & 2.62E-22 & 3.14E-20 & 6.49E-19 & 4.37E-19 & 2.94E-19 & 1.83E-19 & 9.51E-20 \\
1.58E+05 & 4.48E-24 & 2.26E-22 & 1.13E-21 & 2.03E-22 & 3.84E-21 & 7.81E-20 & 6.72E-19 & 5.15E-19 & 3.98E-19 & 2.73E-19 \\
2.51E+05 & 3.31E-24 & 9.86E-23 & 1.68E-21 & 1.71E-21 & 9.10E-22 & 7.95E-21 & 9.78E-20 & 6.87E-19 & 6.43E-19 & 4.86E-19 \\
3.98E+05 & 2.72E-24 & 5.31E-23 & 6.76E-22 & 3.54E-21 & 1.87E-21 & 1.73E-21 & 1.01E-20 & 9.69E-20 & 6.78E-19 & 7.34E-19 \\
6.31E+05 & 2.55E-24 & 3.32E-23 & 3.09E-22 & 2.16E-21 & 5.26E-21 & 2.60E-21 & 2.57E-21 & 1.10E-20 & 7.55E-20 & 3.88E-19 \\
1.00E+06 & 2.64E-24 & 2.42E-23 & 1.74E-22 & 9.46E-22 & 4.32E-21 & 7.36E-21 & 4.37E-21 & 3.69E-21 & 1.13E-20 & 3.74E-20 \\
1.58E+06 & 2.91E-24 & 2.07E-23 & 1.11E-22 & 5.01E-22 & 1.97E-21 & 6.16E-21 & 1.08E-20 & 7.69E-21 & 6.14E-21 & 8.81E-21 \\
2.51E+06 & 3.35E-24 & 2.00E-23 & 8.35E-23 & 3.07E-22 & 1.04E-21 & 2.96E-21 & 8.04E-21 & 1.44E-20 & 1.33E-20 & 9.50E-21 \\
3.98E+06 & 3.95E-24 & 2.10E-23 & 7.25E-23 & 2.18E-22 & 6.27E-22 & 1.62E-21 & 4.01E-21 & 9.23E-21 & 1.70E-20 & 1.94E-20 \\
6.31E+06 & 4.74E-24 & 2.32E-23 & 7.02E-23 & 1.81E-22 & 4.41E-22 & 9.99E-22 & 2.26E-21 & 4.83E-21 & 9.70E-21 & 1.76E-20 \\
1.00E+07 & 5.75E-24 & 2.66E-23 & 7.35E-23 & 1.69E-22 & 3.60E-22 & 7.24E-22 & 1.45E-21 & 2.84E-21 & 5.38E-21 & 9.76E-21 \\
1.58E+07 & 7.08E-24 & 3.15E-23 & 8.16E-23 & 1.73E-22 & 3.34E-22 & 6.08E-22 & 1.09E-21 & 1.93E-21 & 3.35E-21 & 5.68E-21 \\
2.51E+07 & 8.76E-24 & 3.79E-23 & 9.41E-23 & 1.89E-22 & 3.40E-22 & 5.76E-22 & 9.49E-22 & 1.54E-21 & 2.44E-21 & 3.82E-21 \\
3.98E+07 & 1.09E-23 & 4.61E-23 & 1.11E-22 & 2.15E-22 & 3.71E-22 & 5.96E-22 & 9.23E-22 & 1.39E-21 & 2.07E-21 & 3.02E-21 \\
6.31E+07 & 1.35E-23 & 5.65E-23 & 1.34E-22 & 2.52E-22 & 4.22E-22 & 6.55E-22 & 9.73E-22 & 1.40E-21 & 1.97E-21 & 2.73E-21 \\
1.00E+08 & 1.69E-23 & 6.97E-23 & 1.63E-22 & 3.02E-22 & 4.94E-22 & 7.50E-22 & 1.08E-21 & 1.51E-21 & 2.06E-21 & 2.74E-21 \\
1.58E+08 & 2.12E-23 & 8.67E-23 & 2.01E-22 & 3.68E-22 & 5.94E-22 & 8.88E-22 & 1.26E-21 & 1.72E-21 & 2.28E-21 & 2.97E-21 \\
2.51E+08 & 2.67E-23 & 1.08E-22 & 2.48E-22 & 4.52E-22 & 7.23E-22 & 1.07E-21 & 1.50E-21 & 2.02E-21 & 2.64E-21 & 3.38E-21 \\
3.98E+08 & 3.37E-23 & 1.35E-22 & 3.08E-22 & 5.58E-22 & 8.88E-22 & 1.30E-21 & 1.81E-21 & 2.42E-21 & 3.13E-21 & 3.96E-21 \\
6.31E+08 & 4.30E-23 & 1.69E-22 & 3.84E-22 & 6.92E-22 & 1.10E-21 & 1.60E-21 & 2.21E-21 & 2.94E-21 & 3.78E-21 & 4.74E-21 \\
1.00E+09 & 5.56E-23 & 2.11E-22 & 4.79E-22 & 8.60E-22 & 1.36E-21 & 1.98E-21 & 2.72E-21 & 3.59E-21 & 4.60E-21 & 5.76E-21 \\
\end{tabular}%
  \label{tab:coolingHtoNe}%
\end{table}%

%\clearpage

\begin{table}
  \centering
  \caption{Element specific cooling [erg cm$^3$ s$^{-1}$ for the lightest 30 elements - continued}
    \begin{tabular}{ccccccccccc}
\hline
\hline
    Te (K)   & Sodium & Magnesium & Aluminium & Silicon & Phosphorus & Sulphur & Chlorine & Argon & Potassium & Calcium \\
\hline
1.00E+04 & 9.48E-23 & 1.85E-20 & 1.08E-20 & 1.36E-20 & 5.56E-21 & 2.03E-20 & 4.34E-21 & 8.91E-24 & 2.96E-23 & 3.31E-19 \\
1.58E+04 & 1.64E-23 & 1.29E-19 & 7.46E-20 & 3.38E-20 & 2.44E-20 & 1.13E-19 & 5.81E-20 & 5.34E-21 & 8.09E-24 & 2.25E-19 \\
2.51E+04 & 4.98E-24 & 1.80E-20 & 2.81E-19 & 1.47E-19 & 7.30E-20 & 1.61E-19 & 8.99E-20 & 1.31E-20 & 5.99E-23 & 1.14E-20 \\
3.98E+04 & 3.62E-23 & 1.34E-21 & 6.30E-19 & 4.74E-19 & 2.13E-19 & 2.42E-19 & 1.68E-19 & 6.44E-20 & 1.52E-20 & 2.97E-21 \\
6.31E+04 & 1.98E-21 & 2.43E-22 & 3.72E-20 & 9.23E-19 & 6.48E-19 & 7.50E-19 & 1.13E-19 & 2.45E-19 & 1.33E-19 & 4.78E-20 \\
1.00E+05 & 1.81E-20 & 2.22E-21 & 4.10E-21 & 8.18E-19 & 1.46E-18 & 1.55E-18 & 1.38E-19 & 8.28E-19 & 6.73E-19 & 2.57E-19 \\
1.58E+05 & 9.69E-20 & 3.26E-20 & 4.63E-21 & 2.02E-20 & 2.44E-19 & 1.61E-18 & 7.12E-19 & 1.75E-18 & 1.16E-18 & 9.54E-19 \\
2.51E+05 & 2.85E-19 & 1.61E-19 & 5.27E-20 & 1.60E-20 & 2.46E-20 & 7.22E-20 & 1.39E-18 & 2.53E-18 & 1.04E-18 & 1.90E-18 \\
3.98E+05 & 4.85E-19 & 4.14E-19 & 1.94E-19 & 9.84E-20 & 3.35E-20 & 1.09E-20 & 1.38E-19 & 8.43E-19 & 1.22E-18 & 2.53E-18 \\
6.31E+05 & 6.33E-19 & 6.75E-19 & 3.75E-19 & 2.99E-19 & 1.26E-19 & 6.88E-20 & 6.20E-20 & 1.11E-19 & 3.30E-19 & 1.29E-18 \\
1.00E+06 & 1.82E-19 & 5.38E-19 & 5.33E-19 & 5.45E-19 & 2.54E-19 & 2.38E-19 & 9.17E-20 & 1.12E-19 & 1.10E-19 & 1.93E-19 \\
1.58E+06 & 2.41E-20 & 6.16E-20 & 2.91E-19 & 4.73E-19 & 3.28E-19 & 4.23E-19 & 5.08E-20 & 2.47E-19 & 8.84E-20 & 1.67E-19 \\
2.51E+06 & 1.02E-20 & 1.43E-20 & 3.79E-20 & 7.83E-20 & 2.11E-19 & 4.10E-19 & 2.79E-20 & 4.05E-19 & 4.86E-20 & 3.00E-19 \\
3.98E+06 & 1.64E-20 & 1.33E-20 & 1.52E-20 & 2.02E-20 & 3.71E-20 & 5.69E-20 & 1.17E-20 & 2.83E-19 & 3.38E-20 & 4.21E-19 \\
6.31E+06 & 2.40E-20 & 2.47E-20 & 2.19E-20 & 1.93E-20 & 2.03E-20 & 1.36E-20 & 7.30E-21 & 3.89E-20 & 1.30E-20 & 1.92E-19 \\
1.00E+07 & 1.65E-20 & 2.51E-20 & 3.20E-20 & 3.23E-20 & 2.98E-20 & 2.08E-20 & 1.51E-20 & 1.87E-20 & 1.18E-20 & 3.47E-20 \\
1.58E+07 & 9.51E-21 & 1.53E-20 & 2.35E-20 & 3.22E-20 & 3.86E-20 & 3.76E-20 & 3.37E-20 & 3.11E-20 & 2.45E-20 & 2.67E-20 \\
2.51E+07 & 6.01E-21 & 9.26E-21 & 1.40E-20 & 2.06E-20 & 2.88E-20 & 3.65E-20 & 4.28E-20 & 4.63E-20 & 4.44E-20 & 4.30E-20 \\
3.98E+07 & 4.43E-21 & 6.42E-21 & 9.21E-21 & 1.31E-20 & 1.83E-20 & 2.44E-20 & 3.21E-20 & 4.05E-20 & 4.74E-20 & 5.28E-20 \\
6.31E+07 & 3.78E-21 & 5.17E-21 & 7.03E-21 & 9.51E-21 & 1.28E-20 & 1.65E-20 & 2.16E-20 & 2.78E-20 & 3.49E-20 & 4.27E-20 \\
1.00E+08 & 3.63E-21 & 4.75E-21 & 6.16E-21 & 7.97E-21 & 1.02E-20 & 1.28E-20 & 1.62E-20 & 2.03E-20 & 2.53E-20 & 3.11E-20 \\
1.58E+08 & 3.81E-21 & 4.83E-21 & 6.06E-21 & 7.56E-21 & 9.38E-21 & 1.14E-20 & 1.39E-20 & 1.70E-20 & 2.05E-20 & 2.47E-20 \\
2.51E+08 & 4.25E-21 & 5.27E-21 & 6.47E-21 & 7.87E-21 & 9.51E-21 & 1.13E-20 & 1.35E-20 & 1.60E-20 & 1.89E-20 & 2.22E-20 \\
3.98E+08 & 4.93E-21 & 6.03E-21 & 7.29E-21 & 8.72E-21 & 1.04E-20 & 1.22E-20 & 1.42E-20 & 1.66E-20 & 1.92E-20 & 2.21E-20 \\
6.31E+08 & 5.85E-21 & 7.10E-21 & 8.50E-21 & 1.01E-20 & 1.18E-20 & 1.38E-20 & 1.59E-20 & 1.83E-20 & 2.09E-20 & 2.38E-20 \\
1.00E+09 & 7.06E-21 & 8.52E-21 & 1.01E-20 & 1.19E-20 & 1.39E-20 & 1.61E-20 & 1.85E-20 & 2.11E-20 & 2.39E-20 & 2.70E-20 \\
\end{tabular}%
  \label{tab:coolingNatoCa}%
\end{table}%

\clearpage

\begin{table}
  \centering
  \caption{Element specific cooling [erg cm$^3$ s$^{-1}$ for the lightest 30 elements - continued}
    \begin{tabular}{ccccccccccc}
\hline
\hline
    Te (K)  & Scandium & Titanium & Vanadium & Chromium & Manganese & Iron & Cobalt & Nickel & Copper & Zinc \\
\hline
1.00E+04 & 3.77E-19 & 1.42E-19 & 1.35E-20 & 5.28E-21 & 5.37E-20 & 8.84E-20 & 2.74E-21 & 5.26E-21 & 1.07E-21 & 5.83E-21 \\
1.58E+04 & 5.24E-19 & 4.91E-19 & 7.65E-20 & 5.44E-20 & 3.50E-19 & 2.16E-19 & 3.49E-20 & 8.16E-21 & 2.37E-21 & 1.25E-19 \\
2.51E+04 & 9.27E-20 & 1.39E-19 & 3.57E-20 & 1.17E-19 & 1.85E-19 & 9.70E-20 & 1.97E-20 & 2.77E-20 & 1.26E-20 & 2.53E-19 \\
3.98E+04 & 8.02E-21 & 5.90E-20 & 7.25E-20 & 1.06E-19 & 1.31E-20 & 1.12E-19 & 2.76E-20 & 4.72E-21 & 2.58E-21 & 3.47E-20 \\
6.31E+04 & 1.23E-20 & 1.68E-20 & 4.86E-20 & 1.04E-19 & 1.05E-21 & 1.46E-19 & 1.42E-20 & 6.48E-22 & 5.39E-22 & 5.48E-21 \\
1.00E+05 & 1.58E-19 & 6.73E-20 & 3.53E-20 & 7.22E-20 & 1.53E-21 & 1.55E-19 & 2.07E-21 & 1.03E-21 & 9.20E-22 & 3.57E-21 \\
1.58E+05 & 5.39E-19 & 4.37E-19 & 2.69E-19 & 1.49E-19 & 1.38E-20 & 1.83E-19 & 2.55E-21 & 2.28E-21 & 2.23E-21 & 2.38E-21 \\
2.51E+05 & 3.31E-19 & 8.62E-19 & 9.33E-19 & 6.89E-19 & 3.50E-19 & 4.04E-19 & 1.95E-20 & 6.84E-21 & 6.35E-21 & 5.57E-21 \\
3.98E+05 & 2.50E-19 & 3.72E-19 & 1.07E-18 & 1.42E-18 & 1.27E-18 & 9.57E-19 & 3.45E-19 & 1.74E-20 & 1.52E-20 & 1.38E-20 \\
6.31E+05 & 1.06E-18 & 4.09E-19 & 9.72E-19 & 1.30E-18 & 1.76E-18 & 1.61E-18 & 1.35E-18 & 1.09E-19 & 3.49E-20 & 3.39E-20 \\
1.00E+06 & 4.35E-19 & 9.17E-19 & 1.20E-18 & 1.38E-18 & 8.44E-19 & 2.08E-18 & 2.08E-18 & 9.56E-19 & 7.07E-20 & 7.07E-20 \\
1.58E+06 & 1.73E-19 & 2.49E-19 & 4.11E-19 & 7.39E-19 & 7.28E-19 & 1.86E-18 & 1.59E-18 & 1.52E-18 & 9.60E-20 & 9.92E-20 \\
2.51E+06 & 1.85E-19 & 1.88E-19 & 2.10E-19 & 2.53E-19 & 3.79E-19 & 6.53E-19 & 8.78E-19 & 1.08E-18 & 1.38E-19 & 1.36E-19 \\
3.98E+06 & 2.15E-19 & 2.04E-19 & 2.13E-19 & 2.19E-19 & 2.44E-19 & 3.80E-19 & 3.32E-19 & 4.16E-19 & 6.05E-20 & 1.14E-19 \\
6.31E+06 & 2.19E-19 & 1.98E-19 & 2.02E-19 & 2.01E-19 & 2.19E-19 & 4.41E-19 & 2.83E-19 & 3.02E-19 & 4.85E-20 & 5.07E-20 \\
1.00E+07 & 7.69E-20 & 1.05E-19 & 1.47E-19 & 1.56E-19 & 1.60E-19 & 4.59E-19 & 2.17E-19 & 2.51E-19 & 7.00E-20 & 6.76E-20 \\
1.58E+07 & 4.14E-20 & 4.89E-20 & 6.24E-20 & 7.58E-20 & 9.50E-20 & 2.19E-19 & 1.45E-19 & 1.62E-19 & 7.27E-20 & 8.13E-20 \\
2.51E+07 & 4.67E-20 & 4.66E-20 & 4.80E-20 & 5.11E-20 & 5.67E-20 & 8.91E-20 & 6.99E-20 & 8.97E-20 & 3.86E-20 & 4.75E-20 \\
3.98E+07 & 5.72E-20 & 5.86E-20 & 5.87E-20 & 5.87E-20 & 5.90E-20 & 7.16E-20 & 5.84E-20 & 6.39E-20 & 4.08E-20 & 4.25E-20 \\
6.31E+07 & 5.10E-20 & 5.82E-20 & 6.40E-20 & 6.82E-20 & 7.08E-20 & 7.88E-20 & 7.19E-20 & 7.32E-20 & 6.30E-20 & 6.49E-20 \\
1.00E+08 & 3.74E-20 & 4.54E-20 & 5.36E-20 & 6.12E-20 & 6.85E-20 & 7.77E-20 & 8.01E-20 & 8.47E-20 & 8.45E-20 & 9.11E-20 \\
1.58E+08 & 2.88E-20 & 3.53E-20 & 4.21E-20 & 4.85E-20 & 5.61E-20 & 6.49E-20 & 7.21E-20 & 8.01E-20 & 8.68E-20 & 9.77E-20 \\
2.51E+08 & 2.51E-20 & 3.03E-20 & 3.56E-20 & 4.05E-20 & 4.67E-20 & 5.38E-20 & 6.09E-20 & 6.88E-20 & 7.68E-20 & 8.73E-20 \\
3.98E+08 & 2.48E-20 & 2.91E-20 & 3.34E-20 & 3.76E-20 & 4.27E-20 & 4.85E-20 & 5.46E-20 & 6.14E-20 & 6.85E-20 & 7.70E-20 \\
6.31E+08 & 2.65E-20 & 3.05E-20 & 3.44E-20 & 3.84E-20 & 4.30E-20 & 4.81E-20 & 5.35E-20 & 5.96E-20 & 6.59E-20 & 7.30E-20 \\
1.00E+09 & 3.00E-20 & 3.39E-20 & 3.78E-20 & 4.19E-20 & 4.65E-20 & 5.14E-20 & 5.67E-20 & 6.24E-20 & 6.84E-20 & 7.50E-20 \\
\end{tabular}%
  \label{tab:coolingSctoZn}%
\end{table}%

\clearpage

\label{lastpage}
\clearpage

\end{document}